\documentclass[%
reprint, amsmath,amssymb,
aps,
pra,
]{revtex4-1}
\usepackage{bbold}
\usepackage{centernot,cancel}

\usepackage{rotating}
\usepackage{diagbox}
\usepackage{graphicx}
\usepackage{dcolumn}
\usepackage{bm}

\usepackage{dsfont}
\usepackage[english]{babel}
\usepackage{overpic}
\usepackage{amsmath}

\usepackage{amsfonts}
\usepackage{braket}
\usepackage{amssymb}
\usepackage{xcolor}
\usepackage{makeidx}

\usepackage{graphicx}
\usepackage{kpfonts}
\usepackage{lmodern}
\usepackage[left=2cm,right=2cm,top=2cm,bottom=2cm]{geometry}
\usepackage{hyperref}
\hypersetup{
    colorlinks=true,
    linkcolor=blue,
    filecolor=magenta,      
    urlcolor=blue,
    citecolor=red,
}

\def\Xint#1{\mathchoice
   {\XXint\displaystyle\textstyle{#1}}%
   {\XXint\textstyle\scriptstyle{#1}}%
  {\XXint\scriptstyle\scriptscriptstyle{#1}}%
  {\XXint\scriptscriptstyle\scriptscriptstyle{#1}}%
  \!\int}
\def\XXint#1#2#3{{\setbox0=\hbox{$#1{#2#3}{\int}$}
     \vcenter{\hbox{$#2#3$}}\kern-.5\wd0}}

\def\dashint{\Xint-}

\begin{document}

\author{D. Farina$^{1,2}$, G. De Filippis$^{3,4}$, V. Cataudella$^{3,4}$, M. Polini$^{2}$ and V. Giovannetti$^{5}$}

\affiliation{
$^{1}$NEST, Scuola Normale Superiore, I-56126 Pisa, Italy.\\
$^{2}$Istituto Italiano di Tecnologia, Graphene Labs, Via Morego 30, I-16163 Genova, Italy.\\
$^{3}$SPIN-CNR and Dip. di Fisica - Università di Napoli Federico II, I-80126 Napoli, Italy.\\
$^{4}$INFN, Sezione di Napoli - Complesso Universitario di Monte S. Angelo, I-80126 Napoli, Italy.\\
$^{5}$NEST, Scuola Normale Superiore and Istituto Nanoscienze-CNR, I-56127 Pisa, Italy.}

\date{\today}

\begin{abstract}
Identifying which master equation is preferable for the description of a multipartite open quantum system is not trivial and has led in the recent years to the {local vs. global debate} in the context of Markovian dissipation. 
We treat here a paradigmatic scenario in which the system is composed of two interacting harmonic oscillators A and B, with only A interacting with a thermal bath - collection of other harmonic oscillators - and we study the equilibration process of the system initially in the ground state with the bath finite  temperature.
We show that the completely positive version of the Redfield equation obtained using coarse-grain and an appropriate time-dependent convex mixture of the local and global solutions give rise to the most accurate semigroup approximations of the whole exact system dynamics, i.e. both at short and at long time scales, outperforming the  local and global approaches.
\end{abstract}

\title{Going beyond Local and Global approaches for localized thermal dissipation}
\maketitle

\section{Introduction}
In the rising field of quantum technology \cite{riedel2017european}, considering a quantum system isolated from its surroundings is a non-realistic idealization.
In the majority of the implementations of quantum information algorithms \cite{nielsen-chuang-book} and quantum computation \cite{arute2019supremacy},
the interaction with the environment is detrimental for quantum resources, becoming a crucial ingredient to monitor, with the scope of reducing its effects or with the aim of accounting for it by applying quantum error correction methods.
Interestingly, in more rare cases the environment itself acts as a mediator for the production of quantum correlations into the system~\cite{benatti2003environment}.

Unfortunately
 our ability in accounting for environmental effects is severely limited by  the difficulty of 
keeping track  of  the exact dynamics of the entire system-environment compound: a  problem which is made computationally hard  by the large number of degrees of freedom involved in
the process.
For this reason, effective models for the way the environment acts on the reduced system density matrix have been developed, leading to the master equation (ME) formalism~\cite{lindblad1976generators, gorini1976completely}. 
The lowest level of approximation contemplates the assumption of weak system-environment coupling (Born approximation)
and time-divisibility for the system dynamics (Markov approximation).
This  leads to the Redfield equation \cite{redfield1957theory,breuer2002theory, jeske2013derivation} which regrettably, while being able to capture some important features of the model~\cite{lim2017signatures, Purkayastha2016}, does not ensure positive (and hence completely positive) evolution \cite{gaspard1999slippage, argentieri2014violations, ishizaki2009adequacy, benatti2003nonpositive, wilkie2001dissipation, suarez1992memory, dumcke1979proper, benatti2005open}.
In quantum mechanics, the positivity
 of density matrices -- i.e. the fact that all their eigenvalues are non-negative --
 is an essential  property  imposed by 
  the probabilistic interpretation of the theory~\cite{nielsen-chuang-book}. 
Allowing for mathematical  structures that do not comply with such 
requirement paves the way to a series of inconsistencies that include
\textit{negative probabilities} of measurements outcomes, violation of the uncertainty relation, 
and ultimately the non-contractive character of the underlying dynamics.
Ways to correct or to circumvent the pathology exhibited by the Redfield equation 
typically relay on 
 the full \cite{breuer2002theory} or the partial \cite{schaller2008preservation, cresser2017coarse, seah2018refrigeration,jeske2015bloch, rivas2017refined, farina2019psa} implementation of the secular approximation: a coarse-grain temporal  average of the system dynamics which, performed in conjunction with the above mentioned
Born and Markov approximations,  leads to a more reliable differential equation for the system
density matrix known as the 
 Gorini-Kossakowski-Sudarshan-Lindblad (GKSL) master equation~\cite{lindblad1976generators, gorini1976completely}.

The situation becomes more complicated when the system is composed of two or more interacting subsystems that are locally coupled to possibly independent reservoirs~\cite{cattaneo2019psa}. In this case, a brute force application of a full secular approximation
leads to the so called {\it global} ME, a GKSL equation obtained
under the implicit assumption  that  the environment will perceive the composite system 
as a unique body irrespectively from the local structure of their mutual  interactions.
While formally correct in terms of the positivity and complete positivity requirements and
predicting long term  behaviours  which are thermodynamically consistent, 
the resulting ME  is prone to introduce errors in the short term description of the dynamical 
process.
A suitable alternative  is provided by 
the so called {\it local} ME approach where,
contrarily to the Global ME, 
 each subsystem is assumed to independently interacts with its \textit{own} environment, keeping track of the local nature of the microscopic interaction.
 Despite in certain situations it can imply the breaking of the second law of thermodynamics \cite{levy2014local}, it allows for a more precise description of the short term dynamics of the composite system. A local approach is usually allowed when the subsystems interact weakly between each other \cite{hofer2017markovian, adesso2017loc-vs-glob, rivas2010markovian}. As well as the Global ME, the local ME can be microscopically derived \cite{hofer2017markovian} and is in GKSL form.
Notably, such master equation has recently acquired full dignity showing that it 
exactly  
describes the dynamics induced by an \textit{engineered} bath schematized by a collisional model
\cite{de2018reconciliation}. 
Furthermore, even under a more conventional description of the environment, 
thermodynamics inconsistencies only 
occur  at the order of approximation where the local approach is not guaranteed to be valid and, eventually, it is possible to completely cure such inconsistencies by implementing a perturbative treatment   
around the local approximation~\cite{trushechkin2016perturbative}.

The scope of the present work is to test the effectiveness of different classes of MEs  to describe  the system dynamics, particularly focusing on alternative approaches beyond those adopted in deriving the local and global  MEs  and using as benchmark a model that we are able to solve exactly.
Differently from previous studies \cite{hofer2017markovian,adesso2017loc-vs-glob}, where the focus was on the steady state properties of a bipartite system with each subsystem coupled to a different thermal reservoir, 
we deal with a bipartite system  asymmetrically coupled to a single thermal bath and analyze its whole dynamics including both the transient and asymptotic regime.
More specifically, in our case the system of interest  is composed of two interacting harmonic oscillators A and B, with only A  microscopically coupled with an external bosonic thermal bath
described as a collection of extra harmonic oscillators. 
About the exact dynamics benchmark, the unitary evolution of the joint system+environment compound has been calculated by restricting ourself to exchange interactions and gaussian states~\cite{serafini2017quantum}. 
Our analysis leads to the conclusion that
the completely positive version of the Redfield equation  obtained as described in \cite{farina2019psa} by applying the secular approximation via coarse-grain averaging in a partial and tight way, provides a  semigroup description of the system dynamics that 
outperform both the local and Global ME approaches. 
We also observe that analogous advantages can be obtained by adopting a phenomenological 
description of the system dynamics,  constructed in terms of 
an appropriate time-dependent convex mixture of the local and Global ME solutions.

Despite the selected model has been chosen primarily for its minimal character, possible implementations of the set up we deal with can be found in cavity (or in circuit) quantum electrodynamics. 
An example is the open Dicke model \cite{dicke1954coherence} 
for large enough number of two-level atoms inside the cavity \cite{emary2003chaos} and 
assuming that the interaction of the cavity mode with the radiation field is more relevant than the direct coupling of the radiation field with the atoms.
Alternatively, our bipartite system may directly describe coupled cavities in an array  \cite{hartmann2008quantum} in the instance of two cavities.
About the kind of dynamics we chose, 
it may be of interest 
for ground state storage in quantum computation
\cite{nielsen-chuang-book} or, conversely, for thermal charging tasks \cite{farina2019charger, hovhannisyan2020charging}.

The paper is organized as follows. 
In Sec.~\ref{sec:model} we introduce the  model.
The different approximations are described in Sec~\ref{sec:approximations}.
In Sec.~\ref{DYNAMSC}  we integrate the dynamical evolution under the various approximations and 
present a comparison between the various results. 
In Sec~\ref{sec:conclusions} we draw the conclusions and we discuss possible future developments.
Details on the approximation methods and on the evaluation of the exact dynamics are reported in the~Appendix.

\begin{figure}
\includegraphics[width=.6\columnwidth]{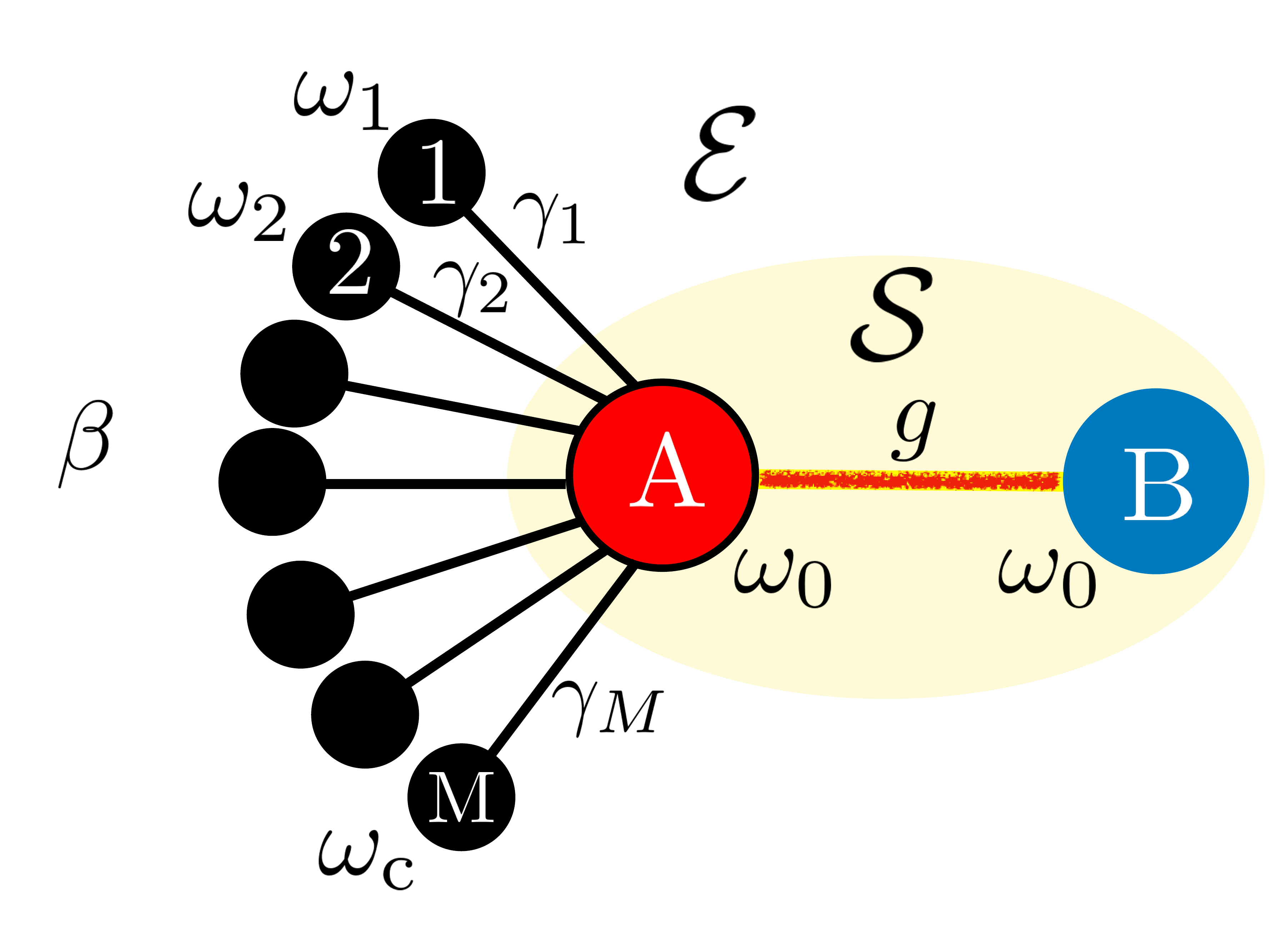}
\caption{Schematic of the  model: the composite system $\mathcal{S}$ is formed of 
two harmonic oscillators A and B of equal frequency  $\omega_0$ which interact via an exchange Hamiltonian coupling characterized by the constant $g$. The subsystem A  is  also coupled  with the modes $k \in \{1, 2, \dots M\}$ of a thermal environment $\mathcal{E}$ at temperature $1/\beta$ (again the interaction is mediated by an exchange Hamiltonian with 
constants $\gamma_k$). 
\label{fig:asymmetric-charging}}
\end{figure}

\section{The model}
\label{sec:model}
The model we consider is schematically described in Fig.~\ref{fig:asymmetric-charging}. 
It consists into a bipartite system $\mathcal{S}$ composed of two resonant bosonic modes A and B of frequency $\omega_0$ and described by the ladder operators $a,a^\dag$ and $b, b^\dag$,
 that interact through an excitation preserving coupling characterized by 
an intensity parameter $g\geq 0$. 
Accordingly, setting $\hbar=1$, the free Hamiltonian of  $\mathcal{S}$ reads 
\begin{eqnarray}
\label{hamiltonian-system}
&H_{\rm S}:=H_{\rm S, 0}+H_{\rm S, g}~,&\\
\nonumber
&H_{\rm S, 0}:=\omega_{\rm A} a^\dagger a + \omega_{\rm B} b^\dagger b ~,
{\rm \qquad with\qquad }\omega_{\rm A}=\omega_{\rm B}:=\omega_0~, &\\
&H_{\rm S, g}:= g(a^\dagger b + {\rm h.c.})~,& \nonumber
\end{eqnarray}
which can also be conveniently expressed as
\begin{eqnarray}
H_{\rm S}&=&\omega_+ \gamma_+^\dag \gamma_+ + \omega_- \gamma_-^\dag \gamma_-~,
\end{eqnarray}
with  
\begin{eqnarray}
\label{eq:eigenmodes}
\omega_\pm &:=& \omega_0\pm g~\;, \qquad 
\gamma_\pm:=\frac{1}{\sqrt{2}} (a\pm b)\;, 
\end{eqnarray}
being, respectively, the associated eigenmode frequencies and operators
~\cite{emary2003chaos}, the last obeying the commutation rules
\begin{eqnarray}
\Big[\gamma_-,\gamma_+\Big]_-=\Big[\gamma_-,\gamma^\dag_+\Big]_-=0~,~ \Big[\gamma_\pm,\gamma^\dag_\pm\Big]_-=1~.
\end{eqnarray}
 Through the exclusive mediation of  subsystem A, we then assume
${\cal S}$   to be connected with an
 external environment ${\cal E}$ formed
of a collection of a large number $M$ of independent bosonic modes,   
no direct coupling being instead allowed between  B and ${\cal E}$. 
Indicating with  $c_k, c_k^\dag$ the ladder operators of the $k$-th mode of 
 ${\cal E}$,
we hence express the full Hamiltonian of the joint system $\mathcal{S}+\mathcal{E}$ as 
\begin{equation}
\label{hamiltonian-decomposition}
H:=H_{\rm S}+H_{\rm E}+H_1\;, 
\end{equation}
 with
\begin{equation}
\label{hamiltonian-bath}
H_{\rm E}:=\sum_{k=1}^M \omega_k c_k^\dagger c_k~, \qquad 
H_1:=\sum_{k=1}^M \gamma_k ( a^\dagger c_k +  {\rm h.c.})\;, 
\end{equation}
being respectively the free Hamiltonian of the environment 
and 
the exchange coupling between A and ${\cal E}$.
More in details, in our analysis
 we shall assume the frequencies $\omega_k$ of the environmental modes to be equally spaced  with a cut-off value $\omega_c> \omega_0$, i.e. 
\begin{equation}
\label{bath-dispersion}
\omega_k := \frac{k}{M} \omega_{\rm c}~, \hspace{2cm} k \in \{1,..., M\}\;,
\end{equation}
and take  the  system-environment coupling constants $\gamma_k$ 
to have the form 
\begin{eqnarray}
\label{eq:gammak-kappaomega0}
\gamma_k:= \sqrt{
\kappa(\omega_0) \left(\frac{\omega_k}{\omega_0}\right)^{\alpha} \frac{\omega_c}{2\pi M}}\;,
\end{eqnarray}
with  $\kappa(\omega_0)$ controlling the effective strength of the interaction between
A and ${\cal E}$. The parameter $\alpha\geq 0$ appearing in Eq.~(\ref{eq:gammak-kappaomega0}) gauges the 
 bath's dispersion relation by imposing the following form for the (rescaled) spectral density of the reservoirs modes~\cite{hofer2017markovian} 
\begin{equation}
\label{eq:decay-rate-function}
\kappa(\omega):=  2 \pi \sum_{k=1}^M \gamma_k^2 \delta(\omega-\omega_k)=
\kappa(\omega_0) \left(\frac{\omega}{\omega_0}\right)^\alpha~ \Theta (\omega_{\rm c}-\omega)\;,
\end{equation}
with $\Theta(x)$ being the Heaviside step function ($\alpha=1$, $\alpha>1$ and $\alpha<1$ being associated to the Ohmic, super-Ohmic, and sub-Ohmic scenarios respectively~\cite{OHM}). 
Finally we shall assume the joint ${\cal S} + {\cal E}$ system to be initialized into a factorized
state 
\begin{eqnarray} \rho_{\rm SE}(0)=\rho_{\rm S}(0) \otimes \rho_{\rm E}(0)\;, \label{ININ} 
\end{eqnarray}  where the
bath is in a thermal state of temperature
 $1/\beta>0$:
\begin{eqnarray}
 \label{iniENV}
 \rho_{\rm E}(0)&:=& \frac{e^{-\beta H_{\rm E}}}{{\rm tr}[e^{-\beta H_{\rm E}}]} = \rho_1(\beta)\otimes \dots \otimes \rho_M(\beta)~,\\ \label{iniENV1}
 \rho_k(\beta)&:=&\frac{e^{-\beta \omega_k c^\dag_k c_k}}{{\rm tr}[e^{-\beta \omega_k c^\dag_k c_k}]}~.
 \end{eqnarray} 

\begin{figure}
\includegraphics[width=.8\linewidth]{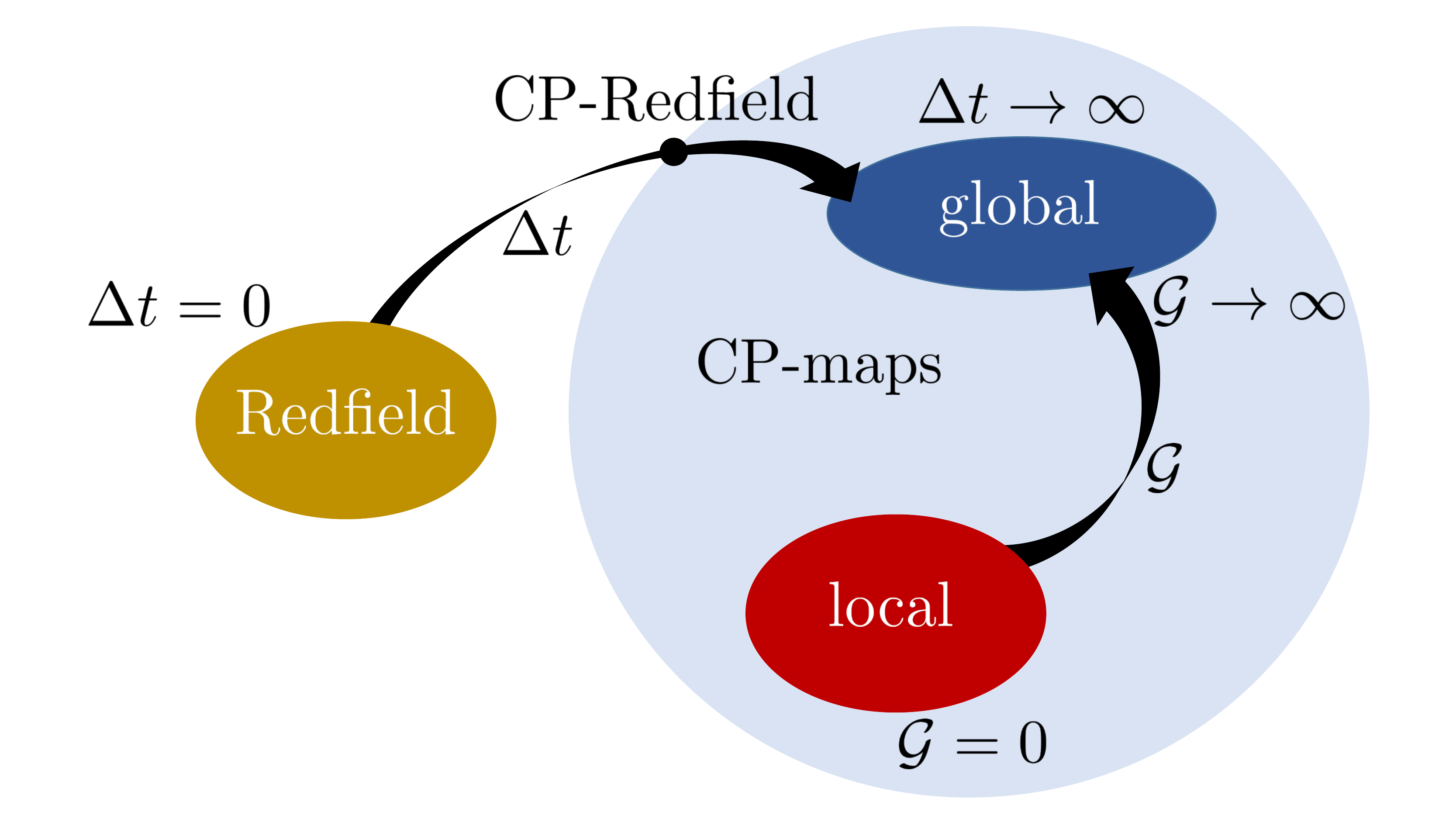}
\caption{Schematic representation of the continuous transitions from the Redfield ME to the Global ME~(\ref{GMEprima})
 passing through the coarse-grained Redfield MEs~(\ref{eq:redfield-coarse-grain-state}), and from the Local ME~(\ref{eq:loc-me-Sch-pict}) to the Global ME using the time-dependent convex mixture~(\ref{eq:convex-comb}).
The dot indicates the completely positive map defined by the CP-Redfield ME obtained by saturating the bound in Eq.~(\ref{eq:positivity-requirement}).}
\label{Fig:Master-eqs}
\end{figure}

\section{Approximated equations for ${\cal S}$}
\label{sec:approximations}
In this section we review the different ME approaches one can use to effectively describe 
the evolution of the system ${\cal S}$ by integrating away  the degrees of freedom 
of  the environment ${\cal E}$. We shall start our presentation by  
introducing the coarse-grained  regularized version of  the Redfield equation~\cite{farina2019psa}, 
which includes  the Global ME as a special case. We then introduce the local ME approach and finally 
discuss the phenomenological approach which employs  convex combinations of local and global 
ME solutions.  Since most of the derivations of the above expressions are 
 discussed in details elsewhere (see e.g.~\cite{breuer2002theory}) here we just give an overview of the methods involved and refer the interested reader to the Appendix~\ref{appendix:approximations} for further details.
\subsection{From CP-Redfield ME to Global ME}
The starting point of this section is the Redfield equation 
 which one obtains by 
expressing the dynamical evolution of the joint system in the interaction picture, and 
enforcing 
the Born and, then, the Markov approximations~\cite{breuer2002theory}.  The Born approximation assumes that the $\mathcal{S}+\mathcal{E}$ coupling is weak in such a way that the state of $\mathcal{E}$ is negligibly influenced by the presence of $\mathcal{S}$, while the Markov approximations assumes invariance of the interaction-picture system state over time-scales of order $\tau_{\rm E}$, the last being the time over which $\mathcal{E}$ loses the information coming from $\mathcal{S}$ and can be estimated from width of the bath correlation functions (see Appendix~\ref{appendix-exact-dynamics}). 

As anticipated in the introductory section, the Redfield equation does not ensure completely positive evolutions and in certain cases neither positive evolution, hence preventing one from framing the obtained
results with the  probabilistic interpretation of quantum mechanics. 
To cure this issue we refer to the version of the partial secular approximation described in Ref.~\cite{farina2019psa}.
Performing a coarse-grain averaging on the Redfield equation in interaction picture over a time interval $\Delta t$ that is much larger than the typical time scale of the system state in interaction picture, is a way to appropriately smooth the non-secular terms responsible of the non-positive  character, even in a tight way.
As schematically pictured in Fig.~\ref{Fig:Master-eqs}, 
by moving the parameter $\Delta t$
 along the interval $[0,\infty[$
the reported 
 technique is also capable to formally connect  the original Redfield equation ($\Delta t=0$) and the full secular approximation ($\Delta t=\infty$) in a continuous
 way. 
Expressed in Schr\"{o}dinger picture, 
 the coarse-grained Redfield equation for the evolution of $\rho_{\rm S}$ for
 fixed coarse-graining time $\Delta t$, 
reads
\begin{eqnarray}
\label{eq:redfield-coarse-grain-state}
&&\dot{\rho_{\rm S}}(t)=
-i \Big[H_{\rm S}+H_{\rm LS}^{(\Delta t)}, \rho_{\rm S}(t)\Big]_-
\\
&& \qquad 
+ \sum_{\sigma , \sigma'=\pm}
S_{\sigma \sigma'}^{(\Delta t)}
\Big\{ 
\gamma^{(1)}_{\sigma \sigma'}\Big(\gamma_{\sigma}^\dag {\rho_{\rm S}}(t) \gamma_{\sigma'}
-\frac{1}{2}\Big[ \gamma_{\sigma'} \gamma_{\sigma}^\dag,{\rho_{\rm S}}(t)\Big]_+\Big)
\nonumber
\\
&&\qquad \qquad +
\gamma^{(2)}_{\sigma' \sigma}\Big( \gamma_{\sigma'}{\rho_{\rm S}}(t) \gamma_{\sigma}^\dag
-\frac{1}{2}\Big[ \gamma_{\sigma}^\dag \gamma_{\sigma'} ,{\rho_{\rm S}}(t)\Big]_+\Big)\Big\},
\nonumber
\end{eqnarray} 
where hereafter we shall use the symbols $\Big[\cdots, \cdots\Big]_{\mp}$ to represents commutator and  anti-commutator relations, where $\gamma_{\pm}
$ are the eigenmode operators of $H_{\rm S}$ introduced in Eq.~(\ref{eq:eigenmodes}), 
and where 
\begin{eqnarray} 
H_{\rm LS}^{(\Delta t)}:=
\sum_{\sigma , \sigma'=\pm} S_{\sigma \sigma'}^{(\Delta t)}
(\eta_{\sigma \sigma'}^{(1)}  +
\eta_{\sigma' \sigma}^{(2)})
\gamma^\dag_{\sigma} \gamma_{\sigma'}~,
&
\end{eqnarray}
being the so called Lamb-shift Hamiltonian correction term.
As indicated by the notation,
 the dependence of Eq.~(\ref{eq:redfield-coarse-grain-state}) 
 upon the coarse-graining time interval $\Delta t$
 is carried out by the tensor $S_{\sigma \sigma'}^{(\Delta t)}$ 
of components 
 \begin{eqnarray}
\label{eq:sinc-sigma-sigma1}
S_{\sigma \sigma'}^{(\Delta t)}
&:=& 
{\rm sinc}\left(\tfrac{(\sigma-\sigma')g\Delta t}{2}\right)
\\
\nonumber
&=&\delta_{\sigma \sigma'} + (1-\delta_{\sigma \sigma'}) \; {\rm sinc}(g \Delta t)~,
\end{eqnarray}
with
${\rm sinc}(x):= \sin(x)/x$ being the cardinal sinus.
The functional dependence of the right-hand-side of
(\ref{eq:redfield-coarse-grain-state}) 
upon the bath temperature
  is instead  carried on by
the tensors
$\gamma^{(i)}_{\sigma \sigma'}$ and $\eta^{(i)}_{\sigma \sigma'}$. 
 Specifically, for 
 $\sigma, \sigma' \in \{+, -\}$  and $i\in\{ 1,2\}$, 
these elements fulfill the constraints
\begin{eqnarray}
\label{eq:gamma-eta-sigma-sigma'}
\gamma^{(i)}_{\sigma \sigma'}:=\frac{\gamma^{(i)}_{\sigma \sigma}+\gamma^{(i)}_{\sigma' \sigma'}}{2}+
i (\eta^{(i)}_{\sigma \sigma}-\eta^{(i)}_{\sigma' \sigma'})~,
\\ \label{eq:gamma-eta-sigma-sigma'1}
\eta^{(i)}_{\sigma \sigma'}:=-i \frac{\gamma^{(i)}_{\sigma \sigma}-\gamma^{(i)}_{\sigma' \sigma'}}{4}+
\frac{\eta^{(i)}_{\sigma \sigma}+\eta^{(i)}_{\sigma' \sigma'}}{2}~,
\end{eqnarray}
 which allows one to express all of them in terms of their diagonal ($\sigma= \sigma'$)
 components 
%
\begin{eqnarray}
\gamma_{\sigma \sigma}^{(1)}&:=&
\frac{1}{2}~ 
\kappa(\omega_\sigma)
~\mathcal{N}(\omega_\sigma)~,
\\
\gamma_{\sigma \sigma}^{(2)}&:=&
\frac{1}{2}~ 
\kappa(\omega_\sigma)
~[1+\mathcal{N}(\omega_\sigma)]~,
\\
\eta^{(1)}_{\sigma \sigma}&:=&
\frac{1}{2}
\dashint_0^\infty d \epsilon ~
\frac{1}{2 \pi}
\frac{\kappa(\epsilon) \mathcal{N}(\epsilon)}{\epsilon-\omega_\sigma}~,\\
\eta^{(2)}_{\sigma \sigma}&:=&
-\frac{1}{2}
\dashint_0^\infty   d \epsilon ~ 
\frac{1}{2 \pi}
\frac{\kappa(\epsilon) [1+\mathcal{N}(\epsilon)]}{\epsilon-\omega_\sigma}\;,
\end{eqnarray}
with  $
\kappa(\omega)
$ the spectral density of the reservoirs 
defined in Eq.~(\ref{eq:decay-rate-function}) and with 
\begin{equation}
\label{eq:bose-factor}
\mathcal{N}(\omega_k):=\mbox{Tr}[  c_k^\dag c_k \rho_k(\beta)] =\frac{1}{e^{\beta \omega_k}-1}\;,
\end{equation}
being the Bose-Einstein factor of the mode $k$ of the thermal bath.

For  $g\Delta t\rightarrow 0$, $S_{\sigma \sigma'}^{(\Delta t)}$  assumes constant  value $1$ for all $\sigma$ and $\sigma'$: this corresponds to the pathological case of the (uncorrected) Redfield equation in which both the diagonal (secular) and the off-diagonal 
(non-secular) $\sigma, \sigma'$
terms  of the right-hand-side of 
Eq.~(\ref{eq:redfield-coarse-grain-state}) 
 contribute   at the same level to the dynamical evolution of $\rho_{\rm S}(t)$
 paving the way to unwanted non-positive effects. 
As $g\Delta t$ increases 
the off-diagonal component  $S_{+-}^{(\Delta t)}= {\rm sinc}(g \Delta t)$  acts as the smoothing factor for the non-secular ($\sigma\neq \sigma'$) part of the ME,
which 
gets progressively depressed as 
 the coarse-grain time interval $\Delta t$ gets comparable or even larger 
 than the inverse of the  energy scale $g$ of the system.
Following Ref.~\cite{farina2019psa}  one can then 
show that the model 
admits a (finite) threshold value for $\Delta t$ above which Eq.~(\ref{eq:redfield-coarse-grain-state}) acquires  
 the explicit  GKSL form that is necessary and sufficient to ensure
complete positivity of the resulting evolution. 
Specifically, as discussed in details  in Appendix~\ref{CPSEC}, such threshold is triggered by the inequality
\begin{eqnarray}
\label{eq:positivity-requirement}
|{S_{+ -}^{(\Delta t)}}|\leq \min_{i \in \{1,2\}}
\sqrt{
\tfrac{
\gamma_{++}^{(i)}\gamma_{--}^{(i)} 
}{
\lvert\gamma_{+, -}^{(i)}\lvert^2
}}\;.
%
\end{eqnarray}
In the following, the  equation (\ref{eq:redfield-coarse-grain-state}) at positivity threshold, i.e. with the choice of ${S_{+ -}^{(\Delta t)}}$ tightly saturating the bound of Eq.~(\ref{eq:positivity-requirement}), will be called CP-Redfield.

\begin{figure}
\begin{overpic}[width=.98\linewidth]{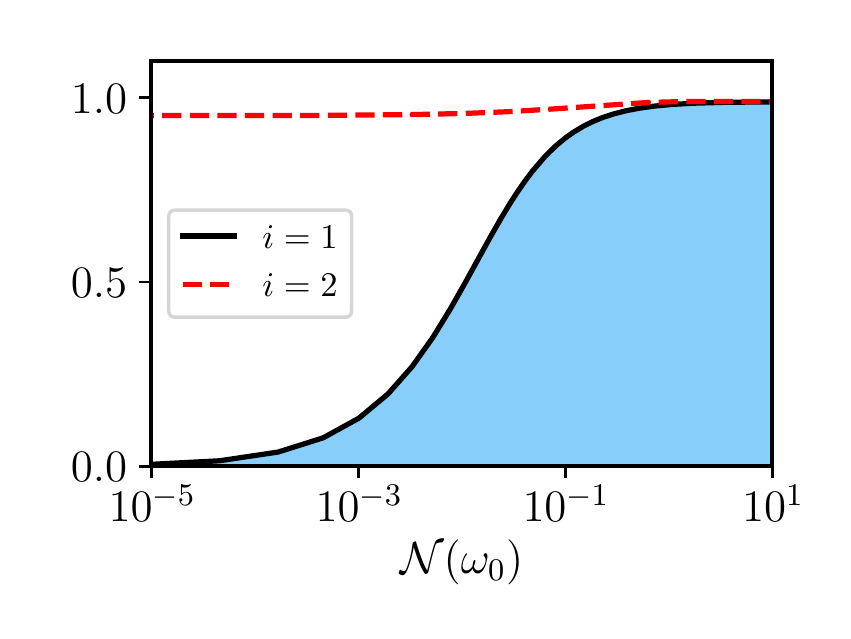}\end{overpic}
\caption{\label{fig:pos-thr}(Color online) 
Plot of the quantities in the right hand side of the inequality (\ref{eq:positivity-requirement})
for i=1 (black full line) and i=2 (red dashed line) as function of 
the bath temperature $1/\beta$ which we parametrize through $\mathcal{N}(\omega_0)= {1}/({e^{\beta \omega_0}-1})$.
The blue region represents the values of $|{S_{+ -}^{(\Delta t)}}|$ which 
satisfy the
 inequality (\ref{eq:positivity-requirement}) ensuring completely positive dynamics of the coarse-grain Redfield equation~(\ref{eq:redfield-coarse-grain-state}).
We chose the parameters $g=0.3 \omega_0$, $\omega_c=3 \omega_0$, and $\alpha=1$ (Ohmic spectral density regime). 
Notice the logarithmic scale on the abscissa.}
\end{figure}

A numerical study of the condition~(\ref{eq:positivity-requirement})
for some selected values of the system parameters is presented
in  
Fig.~\ref{fig:pos-thr}.
This plot makes it clear that the low temperature regime 
($\mathcal{N}(\omega_0)\ll 1$) 
constraints one to take very small values of $|{S_{+ -}^{(\Delta t)}}|$ to guarantee the completely positive character of the evolution \cite{farina2019psa}, while just a tiny correction is needed at high temperatures. These  facts are 
in full agreement with the observation  \cite{suarez1992memory,cheng2005markovian, ishizaki2009adequacy} that the non-positivity character of the Redfield equation is  enhanced at low temperature as a signature of the deviations from 
the Born-Markov assumptions underlying it \cite{hartmann2020accuracy}. We stress that, in this context, non-positivity is originated by the multipartite nature of $\mathcal{S}$: indeed, as $g\rightarrow 0$, the right-hand side of Eq.~(\ref{eq:positivity-requirement}) tends to 1 and consequently the non-positivity of the Redfield ME disappears in this limit.
Notice finally that irrespectively from the value of $g$, Eq.~(\ref{eq:positivity-requirement}) is 
trivially fulfilled 
in the asymptotic 
$g\Delta t
\rightarrow \infty$  limit where $|{S_{+ -}^{(\Delta t)}}|$ approaches the value zero
leading to 
\begin{eqnarray} 
S_{\sigma \sigma'}^{(\infty)}=\delta_{\sigma \sigma'}\;.
\end{eqnarray} 
 This condition identifies 
the full secular approximation
of Eq.~(\ref{eq:redfield-coarse-grain-state}) 
 that transforms such equation into   the Global ME of the model which, 
 for the sake of completeness, we report here in its explicit form 
\begin{eqnarray} \label{GMEprima} 
&&\dot{\rho}_{\rm S}(t)=
-i \Big[H_{\rm S}+
H_{\rm LS}^{\rm (glob)}
, \rho_{\rm S}(t)\Big]_-
\\
&&\; + 
\sum_{\sigma=\pm }\Big\{ 
\frac{1}{2} \kappa(\omega_\sigma)\mathcal{N}(\omega_\sigma)
\Big(\gamma_{\sigma}^\dag  {\rho_{\rm S}}(t) \gamma_{\sigma}
-\frac{1}{2}\Big[ \gamma_{\sigma} \gamma_{\sigma}^\dag, {\rho_{\rm S}}(t)\Big]_+\Big)
\nonumber
\\
&&\; 
+
\frac{1}{2} \kappa(\omega_\sigma)[1+\mathcal{N}(\omega_\sigma)]
\Big( \gamma_{\sigma} {\rho_{\rm S}}(t) \gamma_{\sigma}^\dag
-\frac{1}{2}\Big[ \gamma_{\sigma}^\dag \gamma_{\sigma} ,{\rho_{\rm S}}(t)\Big]_+\Big)\Big\}~,
\nonumber
\end{eqnarray}
with
\begin{eqnarray} 
H_{\rm LS}^{\rm (glob)}&:=& H_{\rm LS}^{(\infty)}=\sum_{\sigma= \pm } 
{\delta \omega_\sigma}  \gamma^\dag_{\sigma} \gamma_{\sigma}\,,\\
{\delta \omega_\sigma} &:=&
\eta_{\sigma \sigma}^{(1)}  
+
\eta_{\sigma \sigma}^{(2)}
=
\frac{1}{4 \pi}
\dashint_0^\infty d\epsilon 
\frac{\kappa(\epsilon)  }{\omega_\sigma-\epsilon}, 
\end{eqnarray}
being  the secular component of the 
Lamb-shift term~\cite{hofer2017markovian}.

\subsection{Local ME}\label{LOCALME} 

The Local ME for ${\cal S}$ is a GKSL equation characterized by 
Lindblad operators which act locally on the mode A.
Explicitly it is given by
\begin{eqnarray}
\label{eq:loc-me-Sch-pict}
&&\dot{\rho_{\rm S}}(t)=
-i\Big[H_{\rm S}+H_{\rm LS}^{\rm (loc)}, \rho_{\rm S}(t)\Big]_-
\\ 
&&\qquad + \kappa(\omega_0)  \mathcal{N}(\omega_0) 
\Big(a^\dagger \rho_{\rm S}(t) a-\frac{1}{2}
\Big[ a a^\dagger, \rho_{\rm S}(t)\Big]_+
\Big) \nonumber \\
&&\qquad  + \kappa(\omega_0) (1+\mathcal{N}(\omega_0)) 
 \Big(a \rho_{\rm S}(t) a^\dagger-\frac{1}{2}
\Big[a^\dagger a, \rho_{\rm S}(t)\Big]_+ 
\Big)\;,  \nonumber 
\end{eqnarray}
with $\kappa(\omega_0)$ and $\mathcal{N}(\omega_0)$
defined as in the previous section and where now the Lamb-shift term is 
expressed as a modification of the local  Hamiltonian of the A mode only, i.e. 
\begin{eqnarray} 
H_{\rm LS}^{\rm (loc)}&:=&\delta \omega_{\rm A}~ a^\dagger a\,,\\
\delta \omega_{\rm A}&:=& \frac{1}{2 \pi}
\dashint_0^\infty d \omega
 \frac{ \kappa(\omega)}{\omega_0-\omega}. 
\end{eqnarray}
 Effectively Eq.~(\ref{eq:loc-me-Sch-pict}) can be obtained 
 starting  from a  Hamiltonian model for the global system ${\cal S}+{\cal E}$
 where one initially completely
neglects the presence of the B mode, enforces the  same approximations
 that leads one to 
  (\ref{GMEprima}) (i.e.
the Born, Markov, and full secular approximation),
and finally introduces B and its coupling with A 
as an additive Hamiltonian contribution in the resulting expression. 
More formally as shown e.g. in Ref.~\cite{hofer2017markovian}, Eq.~(\ref{eq:loc-me-Sch-pict})
can be derived in the  weak internal coupling limit $g \tau_{\rm E}\ll 1
$ ($\tau_{\rm E}$ being the bath memory time scale, see Appendix \ref{appendix-exact-dynamics} for details) which allows one to treat the interaction between A and B as a perturbative correction with respect to the direct A-${\cal E}$ coupling -- see Appendix~\ref{appendix:approximations} for
more on this.

\subsection{Convex mixing  of Local and Global solutions}\label{SEC:CONVMIX} 
As we shall explicitly see in the next section (see Eq.~(\ref{eq:gibbsian-HS})), the main advantage offered by the Global ME~(\ref{GMEprima}) is that it provides
an accurate description of the steady state of ${\cal S}$  at least in the infinitesimally small ${\cal S}+{\cal E}$ 
 coupling regime where  on pure thermodynamic considerations one expects
 independent thermalization of the eigenmodes $\gamma_\pm$ of the system.
On the contrary  the steady state predicted by the Local ME (\ref{eq:loc-me-Sch-pict}) is wrong (even if increasingly accurate as $g/\omega_0 \rightarrow 0$) because it implies the thermalization of the subsystems A and B regardless of the presence of the internal coupling $H_{\rm S, g }$.
Conversely, the Local ME has the quality to predict Rabi oscillations between A and B at shorter time scales, that are completely neglected when adopting the Global ME.

In view of these observations 
a reasonable way of keeping local effects during the transient still maintaining an accurate steady state solution is to adopt an appropriate phenomenological \textit{ansatz}
describing the evolution of ${\cal S}$ in terms of quantum trajectories that interpolate between the solutions $\rho^{(\rm glob)}_{\rm S}(t)$ and $\rho^{(\rm loc)}_{\rm S}(t)$  of the Global and Local ME, see
Fig.~\ref{Fig:Master-eqs}.
The simplest of these construction is provided by the following 
time-dependent mixture
\begin{equation}
\label{eq:convex-comb}
\rho^{(\rm mix)}_{\rm S}(t):= 
e^{- \mathcal{G} t} \rho^{(\rm loc)}_{\rm S}(t) + \left(1-e^{- \mathcal{G} t}\right) 
\rho^{(\rm glob)}_{\rm S}(t)\;.
\end{equation}
In this expression $\mathcal{G}>0$ is an effective rate, whose inverse fix the time scale of the problem that determines when global thermalization effect start dominating the
system dynamics. 
Accordingly Eq.~(\ref{eq:convex-comb}) 
allows us to keep local effects for \textit{short} time scales $t\lesssim \mathcal{G}^{-1}$ and the correct thermalization of the eigenmodes of the system at longer time scales $t\gg \mathcal{G}^{-1}$.
The above formula can be interpreted as follows: the environment needs a finite amount of time to become aware of the presence of the part B because of its short time correlations (Markovian hypothesis). 
The specific value of  $\mathcal{G}$ is a free variable in this model and works as a fitting parameter: its value can be even estimated  quite roughly because of the relatively large time interval at intermediate time scales where the Global and Local approximations look alike (more on this later). 
It is finally worth observing that from the complete positivity properties of both the solutions
of the Global and Local ME, it follows that (\ref{eq:convex-comb}) 
also fulfills such requirement (indeed convex combinations of completely positive
transformations are also completely positive). On the contrary 
at variance with the original expressions~(\ref{GMEprima}) and (\ref{eq:loc-me-Sch-pict}), as well as the CP-Redfield 
expression~(\ref{eq:redfield-coarse-grain-state}), Eq.~(\ref{eq:convex-comb}) will typically exhibit a non Markovian character and will not  be possible to present it 
in the form of a GKSL differential equation.
This property is a direct consequence of the fact 
that the set of Markovian evolutions is not closed under  convex combinations
~\cite{CONVEX}.

\section{Dynamics}\label{DYNAMSC}

In the study of the approximated equations introduced in the previous section, as well as for their comparison with the 
exact  solution of the ${\cal S}+ {\cal E}$ dynamics,
an important simplification  arises from  the choice we made in fixing the
initial condition of ${\cal E}$.
Indeed thanks to Eqs.~(\ref{iniENV}), (\ref{iniENV1}) 
the  resulting CP-Redfield, Global, and Local MEs, happen to be  Gaussian processes~\cite{serafini2017quantum} which
admit complete characterization only in terms of the first and second moments of the field operators $\gamma_\pm$
 (notice that while the mixture~(\ref{eq:convex-comb}) does not
fit into the set of Gaussian processes -- formally speaking it belongs to the convex-hull of such set --  we can still resort to the above simplification by
 exploiting the fact that $\rho^{(\rm mix)}_{\rm S}(t)$ is explicitly given by the sum of the Global
and Local ME solutions).
Accordingly in studying the dynamics of our approximated schemes we can just focus on 
the functions
$\langle \gamma_\sigma  \rangle(t):=
\mbox{Tr} [ \gamma_\sigma  \rho_{\rm S}(t)]$,
$\langle \gamma_\sigma \gamma_{\sigma'}   \rangle(t):=
\mbox{Tr} [ \gamma_\sigma \gamma_{\sigma'} \rho_{\rm S}(t)]$, and $\langle \gamma_\sigma^{\dag} \gamma_{\sigma'}   \rangle(t):=
\mbox{Tr} [ \gamma_\sigma^{\dag}  \gamma_{\sigma'} \rho_{\rm S}(t)]$ whose temporal dependence can be determined
by solving a restricted set of coupled linear differential equations.
 We also observe that
since the full Hamiltonian (\ref{hamiltonian-decomposition}) conserves the total number of excitations in the ${\cal S} + {\cal E}$ model,  coupling between excitations  conserving and  non-conserving  moments are prevented  \cite{cattaneo2019simmetry} yielding further
 simplification in the analysis.

Having clarified these points, in what follows we shall focus on the special case where 
the input state of ${\cal S}$ is fixed assuming that 
 both A and B are initialized  in the  ground states of their local Hamiltonians,
i.e. 
\begin{eqnarray}  
\label{eq:in-cond-universe}
  \rho_{\rm S}(0)&=&\ket{0}_{\rm A}\bra{0} \otimes\ket{0}_{\rm B}\bra{0} ~,
 \label{eq:in-cond-system}
\end{eqnarray} 
with $|{0}\rangle$ representing the zero Fock state of the corresponding mode. 
Under these conditions the input state is Gaussian~\cite{serafini2017quantum} 
and,
evolved under CP-Redfield, Global, Local and
the exact dynamics, will remain Gaussian at all time. Furthermore 
all the first order moments and all the non-excitation-conserving second order
terms exactly nullify, i.e.
\begin{eqnarray}
\langle \gamma_\sigma  \rangle(t) =0 \;, \qquad \langle \gamma_\sigma \gamma_{\sigma'}   \rangle(t) =0 \;, 
\end{eqnarray} 
 leaving only a restricted set of equations to be explicitly integrated. 
In particular, for the case of the coarse-grained Redfield equation
(\ref{eq:redfield-coarse-grain-state}) we get 
\begin{widetext} 
\begin{eqnarray}
\label{eqs-Redf-sinc}
\frac{d}{dt}\langle \gamma_{+}^\dag \gamma_{+} \rangle(t)&=&
-\frac{1}{2}\kappa(\omega_+)[\langle \gamma_{+}^\dag \gamma_{+} \rangle(t)-\mathcal{N}(\omega_+)]
\\
\nonumber
&&+ \;S_{+-}^{(\Delta t)}\times \left\{2~ {\rm Im }\left((\eta^{(1)}_{+-}+\eta^{(2)}_{-+}) \langle \gamma_{-} \gamma_{+}^\dag \rangle(t) \right)
+
 {\rm Re }\left[(\gamma^{(1)}_{+-}-\gamma^{(2)}_{-+}) \langle \gamma_{-} \gamma_{+}^\dag \rangle(t)\right]\right\}
 ~,
\\
\nonumber
\\
\nonumber
%
\frac{d}{dt}\langle \gamma_{-}^\dag \gamma_{-} \rangle(t)&=&
-\frac{1}{2}\kappa(\omega_-)[\langle \gamma_{-}^\dag \gamma_{-} \rangle(t)-\mathcal{N}(\omega_-)]
\\ \nonumber
&&+ \; S_{+-}^{(\Delta t)}\times \left\{-2~ {\rm Im} \left((\eta^{(1)}_{+-}+\eta^{(2)}_{-+}) \langle \gamma_{-} \gamma_{+}^\dag \rangle(t)\right)
+
%
{\rm   Re }\left[(\gamma^{(1)}_{+-}-\gamma^{(2)}_{-+}) \langle \gamma_{-} \gamma_{+}^\dag \rangle(t)\right]
\right\}
~,
\\
\nonumber
\\
%
\nonumber
\frac{d}{dt}\langle \gamma_{-} \gamma_{+}^\dag \rangle(t)&=&
\{i(\omega_+ + \delta \omega_+-\omega_-  -  \delta \omega_-)-\frac{1}{4}[\kappa(\omega_+)+\kappa(\omega_-)]\}\langle \gamma_{-} \gamma_{+}^\dag \rangle(t)\\ \nonumber
&&+ \; S_{+-}^{(\Delta t)}
\times 
\left\{
i (\eta^{(1)}_{-+}+\eta^{(2)}_{+-})   
\left[\langle \gamma_{-}^\dag \gamma_{-} \rangle(t)-
\langle \gamma_{+}^\dag \gamma_{+} \rangle(t) \right]+
%
\gamma^{(1)}_{-+}+\frac{1}{2} (\gamma^{(1)}_{-+}-\gamma^{(2)}_{+-}) 
\left[\langle \gamma_{-}^\dag \gamma_{-} \rangle(t)+ \langle \gamma_{+}^\dag \gamma_{+} \rangle(t)\right] 
 \right\}~,
\end{eqnarray}
\end{widetext}
with
initial values 
\begin{eqnarray}
\label{in-cond-p-m}
\langle \gamma_+^\dag \gamma_+ \rangle(0)&=&\langle \gamma_-^\dag \gamma_- \rangle(0)=
\langle \gamma_- \gamma_+^\dag \rangle(0)=0~,
\end{eqnarray}
imposed by  (\ref{eq:in-cond-system}).
In particular in the case of full secular approximation ($S_{+-}^{(\Delta t)}=0$) the above set of equations become
\begin{eqnarray}
\label{eqs-GLOBAL}
\frac{d}{dt}\langle \gamma_{+}^\dag \gamma_{+} \rangle(t)&=&
-\frac{1}{2}\kappa(\omega_+)[\langle \gamma_{+}^\dag \gamma_{+} \rangle(t)-\mathcal{N}(\omega_+)]\;, 
\\ \nonumber 
\frac{d}{dt}\langle \gamma_{-}^\dag \gamma_{-} \rangle(t)&=&
-\frac{1}{2}\kappa(\omega_-)[\langle \gamma_{-}^\dag \gamma_{-} \rangle(t)-\mathcal{N}(\omega_-)]\;, 
\\ \nonumber 
%
\frac{d}{dt}\langle \gamma_{-} \gamma_{+}^\dag \rangle(t)&=&
\{i(\omega_+ + \delta \omega_+-\omega_-  -  \delta \omega_-)\\ \nonumber 
&& -\frac{1}{4}[\kappa(\omega_+)+\kappa(\omega_-)]\}\langle \gamma_{-} \gamma_{+}^\dag \rangle(t)\;,
\end{eqnarray}
which yield the evolution of the moments for the  Global~ME~(\ref{GMEprima}). 
Similar considerations hold true for  the Local~ME~(\ref{eq:loc-me-Sch-pict}). In this case 
following Refs.~\cite{hofer2017markovian, farina2019charger} we get 
\begin{widetext} 
\begin{eqnarray}
\label{loc-eqs-pm}
\frac{d}{dt}\langle \gamma_+^\dag \gamma_+ \rangle (t) &=& -\frac{1}{2} \kappa(\omega_0) [\langle \gamma_+^\dag \gamma_+ \rangle (t) - \mathcal{N}(\omega_0)+  {\rm Re} \langle \gamma_- \gamma_+^\dag \rangle(t)]  
+ \delta \omega_{\rm A} ~ {\rm Im} \langle \gamma_- \gamma_+^\dag \rangle(t)      ~, \\
\frac{d}{dt}\langle \gamma_-^\dag \gamma_- \rangle (t) &=& 
-\frac{1}{2} \kappa(\omega_0) [\langle \gamma_-^\dag \gamma_- \rangle(t) - \mathcal{N}(\omega_0)
+  {\rm Re} \langle \gamma_- \gamma_+^\dag \rangle(t)] 
- \delta \omega_{\rm A} ~ {\rm Im} \langle \gamma_- \gamma_+^\dag \rangle(t)  
     ~, \nonumber \\
\frac{d}{dt}\langle \gamma_- \gamma_+^\dag \rangle (t) &=& [i 2 g - \frac{1}{2} \kappa(\omega_0)] \langle \gamma_- \gamma_+^\dag \rangle(t) + 
\frac{ \kappa(\omega_0) }{2}\{\mathcal{N}(\omega_0)
-\frac{1}{2}   [\langle \gamma_+^\dag \gamma_+ \rangle (t) + \langle \gamma_-^\dag \gamma_- \rangle(t)]\}
+i ~\frac{\delta \omega_{\rm A}}{2}
   [ \langle \gamma_-^\dag \gamma_- \rangle(t)- \langle \gamma_+^\dag \gamma_+ \rangle (t)]    
    ~,
     \nonumber 
\end{eqnarray}
\end{widetext}
which, for a direct 
 comparison with Eq.~(\ref{eqs-GLOBAL}),
  we 
 express here in terms of the eigenmodes $\gamma_{\pm}$.

\subsection{Evolution of the second moments} \label{SEC:MOME}  
\begin{figure*}
\begin{overpic}[width=0.4\linewidth]{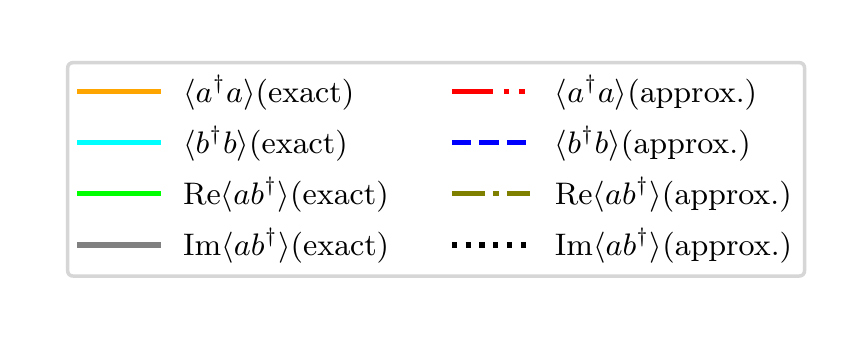}\put(30,73){}\end{overpic}\\
\vspace{.5cm}
\begin{overpic}[width=0.4\linewidth]{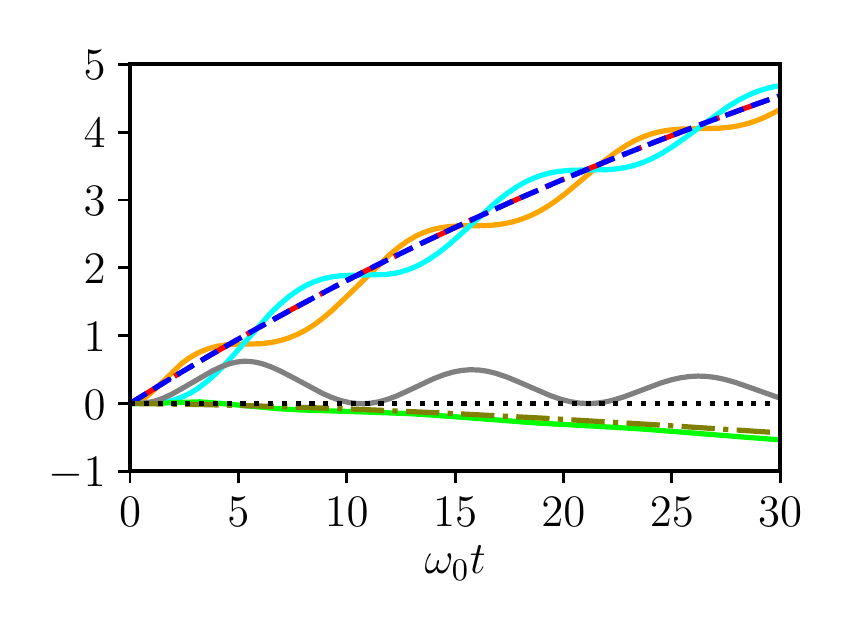}\put(30,73){(a)}\end{overpic}
\begin{overpic}[width=0.4\linewidth]{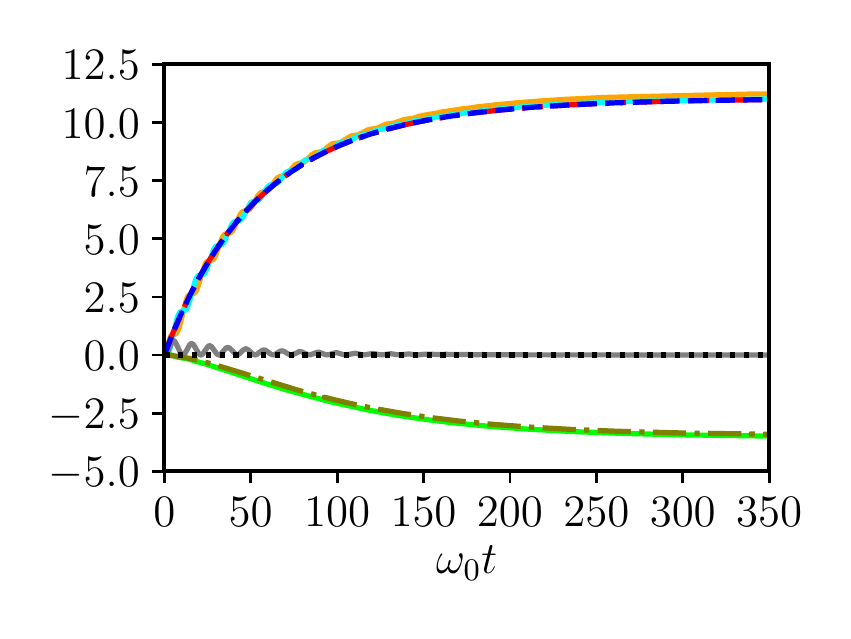}\put(30,73){}\put(-8,75){global}\end{overpic}
\\ \vspace{.5cm}
\begin{overpic}[width=0.4\linewidth]{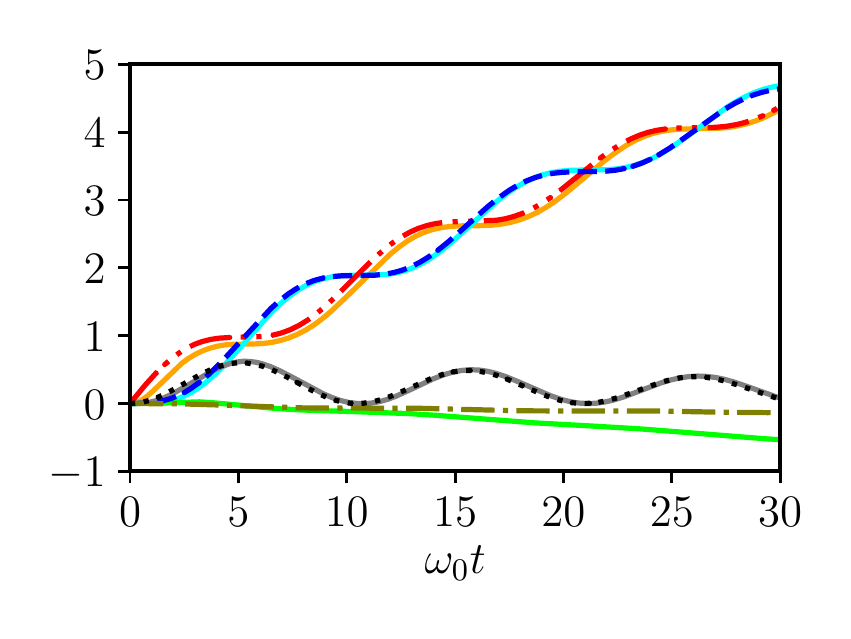}\put(30,73){(b)}\end{overpic}
\begin{overpic}[width=0.4\linewidth]{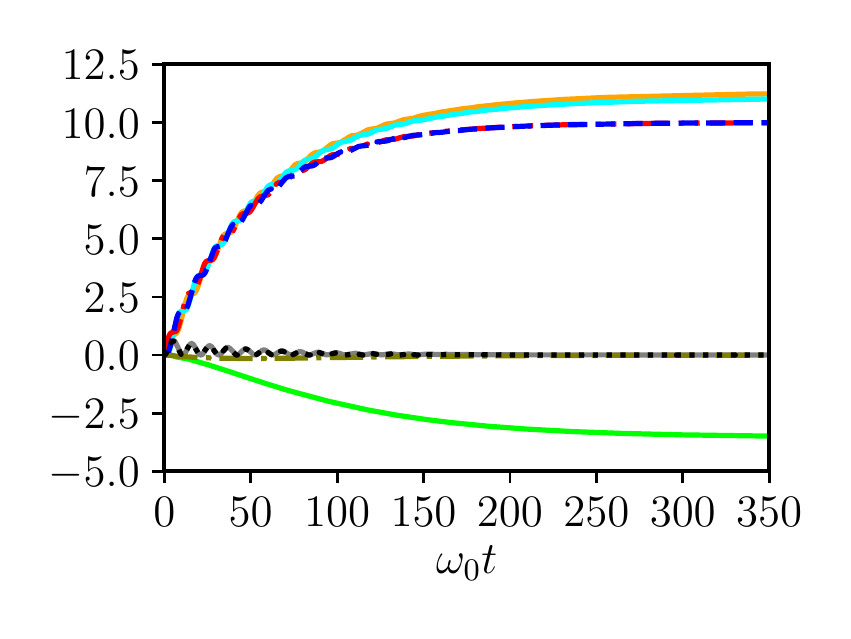}\put(30,73){}\put(-8,75){local}\end{overpic}
\\ \vspace{.5cm}
\begin{overpic}[width=0.4\linewidth]{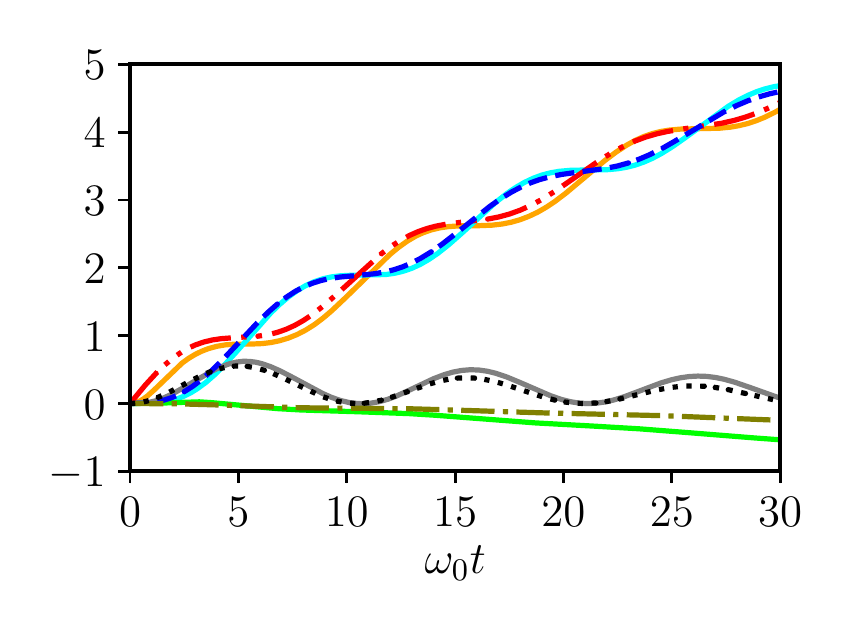}\put(30,73){(c)}\end{overpic}
\begin{overpic}[width=0.4\linewidth]{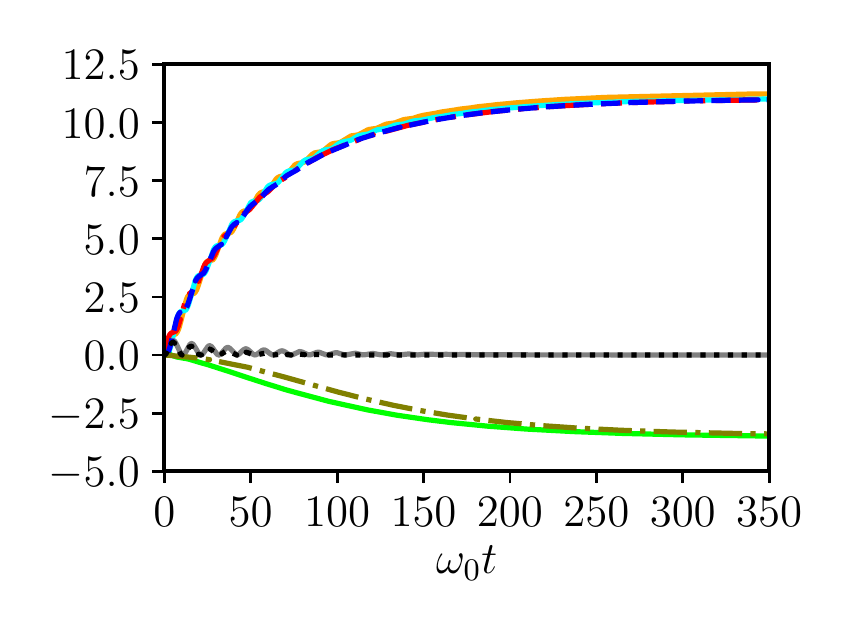}\put(30,73){}\put(-25,75){convex mixture}\end{overpic}
\caption{\label{fig:moms-glob-loc}
(Color online) 
Second order moments evaluated using the Global ME (a), the local ME (b), the convex mixture 
of Eq.~(\ref{eq:convex-comb}) with $\mathcal{G}=0.4 \kappa(\omega_0)$
(c),  
compared with the ones predicted by the exact dynamics. 
As indicated by the legend continuous lines in the plots represent
the quantities computed by solving the exact ${\cal S} + {\cal E}$ Hamiltonian
model~(\ref{hamiltonian-decomposition}); dotted and dashed lines instead
refer to the approximated solutions associated with Global, Local and Mixed 
approaches. 
Each panel contains two plots corresponding each to shorter (left) and longer (right) time scales.
As clear from the right plot of panel (a), the Global ME approach provides a pretty good agreement 
with the exact solutions at large time scales, while fails in the short time domain.
Exactly the opposite occurs for the Local ME approach presented in panel (b):
here a good agreement with the exact solutions is found in the short time domain (left plot), while differences arise in the large time domain (right plot).
The convex mixture approach (\ref{GAMMAMIX}) finally appears to be able to
maintain a good agreement with the exact results at times.  In all the plots we used    $\mathcal{N}(\omega_0)=10$ (corresponding to $1/\beta \approx10.5 \omega_0$),
$g=0.3 \omega_0$,  $\kappa(\omega_0)=0.04 \omega_0$,  $\omega_c=3 \omega_0$, $\alpha=1$.}
\end{figure*}
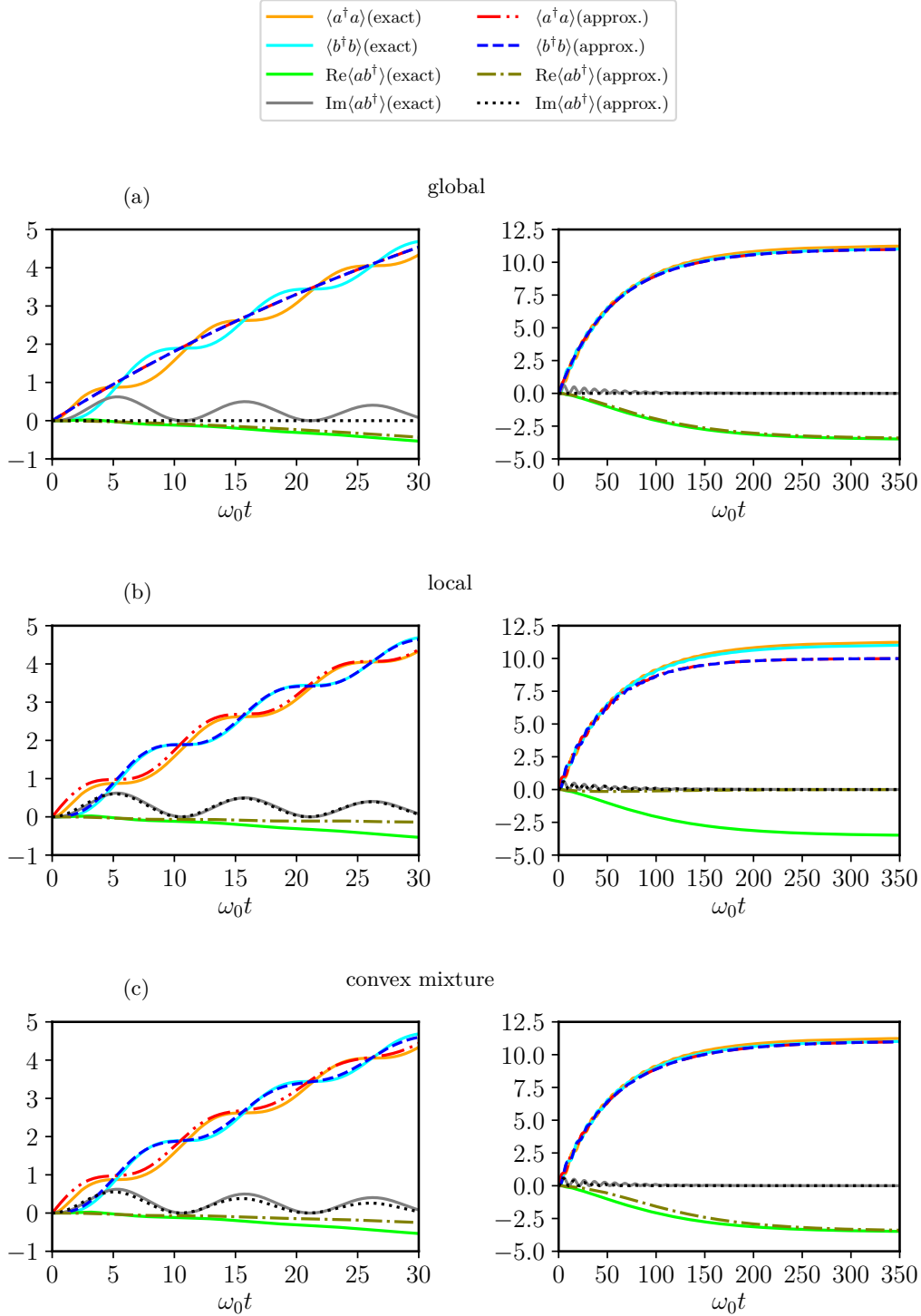
A closer look at Eq.~(\ref{eqs-GLOBAL}) reveals that in this case 
one has that for large enough $t$ we get 
\begin{equation} \label{EQ1} 
\langle \gamma^\dag_{\rm \pm} \gamma_{\rm \pm} \rangle\Big|_   {(\rm glob)} (\infty)= \mathcal{N(\omega_\pm)} \;, \qquad 
\langle \gamma_{\rm -} \gamma^\dag_{\rm +} \rangle\Big|_{(\rm glob)} (\infty)=0\;.
\end{equation}
This enlightens the fact that,
as anticipated at the beginning of Sec.~\ref{SEC:CONVMIX}, the Global ME~(\ref{GMEprima}) 
imposes ${\cal S}$ to asymptotically converge  toward 
 the Gibbs thermal state 
\begin{eqnarray}
\label{eq:gibbsian-HS}
\rho^{(\rm glob)}_{\rm S}(\infty):=\frac{e^{-\beta H_{\rm S}}}{{\rm tr}[ e^{-\beta H_{\rm S}}]}\;,
\end{eqnarray}
in agreement with what one would expect  from purely thermodynamics considerations under 
weak-coupling conditions for the system-environment 
interactions.
On the contrary the steady state predicted by the Local ME is wrong (even if increasingly accurate as $g/\omega_0 \rightarrow 0$) because it implies the thermalization of the subsystems A and B regardless of the presence of the internal coupling $H_{\rm S, g }$. Indeed 
from Eq.~(\ref{loc-eqs-pm}) we get 
\begin{equation} 
\langle \gamma_{\pm}^\dag \gamma_{\pm} \rangle\Big|_{(\rm loc)}(\infty)=\mathcal{N}(\omega_0)~,\qquad \langle \gamma_- \gamma_+^\dag  \rangle\Big|_{(\rm loc)}(\infty)=0
\end{equation} 
or equivalently 
\begin{eqnarray}  \langle a^\dag a \rangle\Big|_{(\rm loc)}(\infty)&=& \langle b^\dag b \rangle\Big|_{(\rm loc)}(\infty)=\mathcal{N}(\omega_0)~,\\
 \langle a b^\dag  \rangle\Big|_{(\rm loc)}(\infty)&=&0\;, \end{eqnarray}
 which identifies 
\begin{eqnarray}
\label{eq:gibbsian-HSlocal}
\rho^{(\rm loc)}_{\rm S}(\infty):=\frac{e^{-\beta H_{\rm S,0}}}{{\rm tr}[ e^{-\beta H_{\rm S,0}}]}\;,
\end{eqnarray}
 as the new fixed point for the dynamical evolution
%
  (see also Appendix \ref{appendix-eigenmodes}).
The discrepancy between the above expressions and
 Eqs.~(\ref{EQ1}), (\ref{eq:gibbsian-HS}) is even accentuated in the low temperature regime   
 $\beta \omega_0 \gg 1$,
where in particular the ratio 
$\mathcal{N}(\omega_-)/\mathcal{N}(\omega_0) \simeq e^{\beta g}$ explodes exponentially.

The situation gets reversed  at shorter time scales. Here 
the Local ME correctly presents coherent energy exchanges between A and B 
which instead the Global approach completely neglects. 
Indeed from Eq.~(\ref{eqs-GLOBAL}) 
 it follows that the Global ME predicts ${\rm Im} [\langle ab^\dag \rangle(t)]=0$, the term  
 being responsible of the Rabi oscillations between A and B.
The Local ME on the contrary 
-- when the Lamb-shift correction can be neglected -- 
gives ${\rm Re}[ \langle ab^\dag \rangle(t)]=0$, the latter  being proportional to the average internal interaction energy $\langle H_{\rm S, g} \rangle$.

The above observations are confirmed by the numerical study we present  in the remaining of the section
(see however also the material presented in Appendix~\ref{APPEX}).
In particular, in panels (a) and (b) of Fig.~\ref{fig:moms-glob-loc} the temporal evolution of the second order moments obtained
by solving Eq.~(\ref{eqs-GLOBAL}) and (\ref{loc-eqs-pm}) are compared with 
 the exact values of the corresponding quantities obtained by numerical
integration of the exact ${\cal S} +{\cal E}$ Hamiltonian model along the lines detailed
in Appendix~\ref{appendix-exact-dynamics}.
In panel (c) of such figure we also  present 
 the results obtained by using the effective model of Sec.~\ref{SEC:CONVMIX}, where according to Eq.~(\ref{eq:convex-comb})
 the expectation values of the relevant quantities are computed as 
\begin{eqnarray}  \label{GAMMAMIX} 
\langle \gamma_\sigma^{\dag} \gamma_{\sigma'}   \rangle\Big|_{(\rm mix)}(t)
&=& e^{- \mathcal{G} t} \langle \gamma_\sigma^{\dag} \gamma_{\sigma'}   \rangle\Big|_{(\rm loc)}(t)\\ \nonumber 
&&  + \left(1-e^{- \mathcal{G} t}\right) 
\langle \gamma_\sigma^{\dag} \gamma_{\sigma'}   \rangle\Big|_{(\rm glob)}(t)\;, 
\end{eqnarray} 
with $\langle \gamma_\sigma^{\dag} \gamma_{\sigma'}   \rangle\Big|_{(\rm loc)}(t)$ and 
$\langle \gamma_\sigma^{\dag} \gamma_{\sigma'}   \rangle\Big|_{(\rm glob)}(t)$ representing  the solutions of
Eq.~(\ref{loc-eqs-pm}) and 
Eq.~(\ref{eqs-GLOBAL}) respectively. 
In our analysis the system  parameters  have been set in order to enforce 
 ${\cal S} + {\cal E}$ weak-coupling conditions
($\omega_0,\omega_\pm \gg \kappa(\omega_0)$)  to make sure that 
that 
the long term prediction 
(\ref{eq:gibbsian-HS}) of 
the Global ME provides a proper description of the system dynamics.
By the same token, the temperature of the bath  has been fixed to be relatively high,
i.e. $1/\beta \approx10.5 \omega_0$, to avoid 
 to enhance 
 correlation effects between the bath and the system which are not included in 
 the Born and Markov approximations needed to derive both the Global and the Local ME~\cite{hovhannisyan2020charging} (a study of the impact of
 low temperature effects on the ${\cal S} + {\cal E}$ correlations is presented in  Appendix~\ref{LOWT}). Finally regarding the value of the phenomenological parameter ${\cal G}$ entering in (\ref{GAMMAMIX}) we set it be equal to $0.4 \kappa(\omega_0)$
finding a relatively good agreement with the exact data at all times.

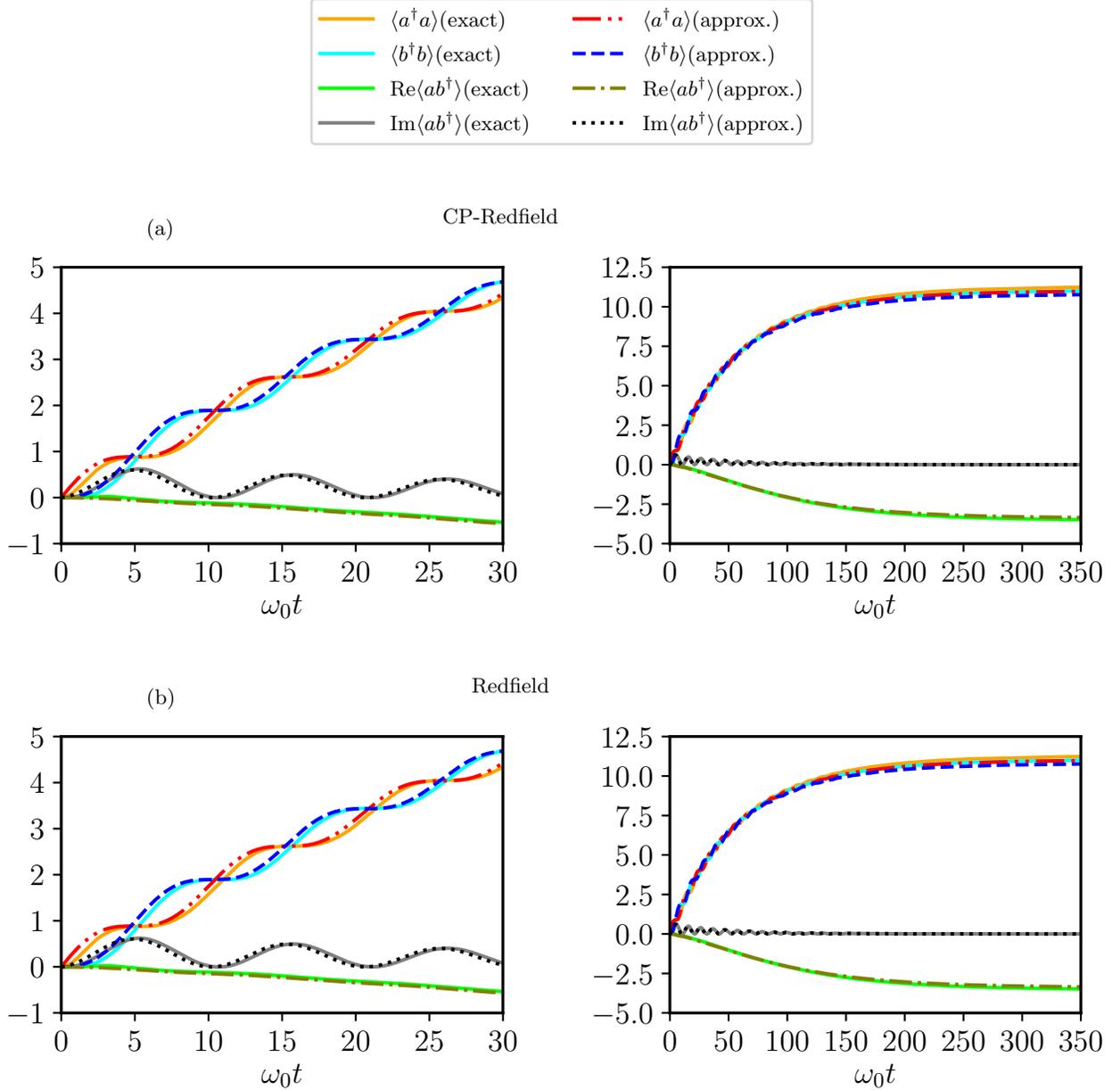
\begin{figure*}
\begin{overpic}[width=0.49\linewidth]{figs/legend}\put(30,73){}\end{overpic}
\\ \vspace{.5cm}
\begin{overpic}[width=0.49\linewidth]{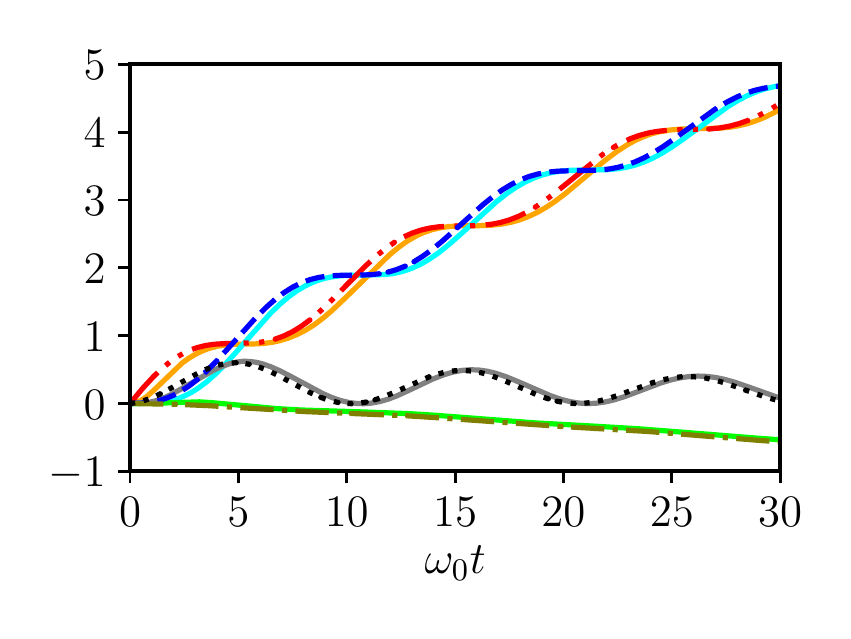}\put(30,73){(a)}\end{overpic}
\begin{overpic}[width=0.49\linewidth]{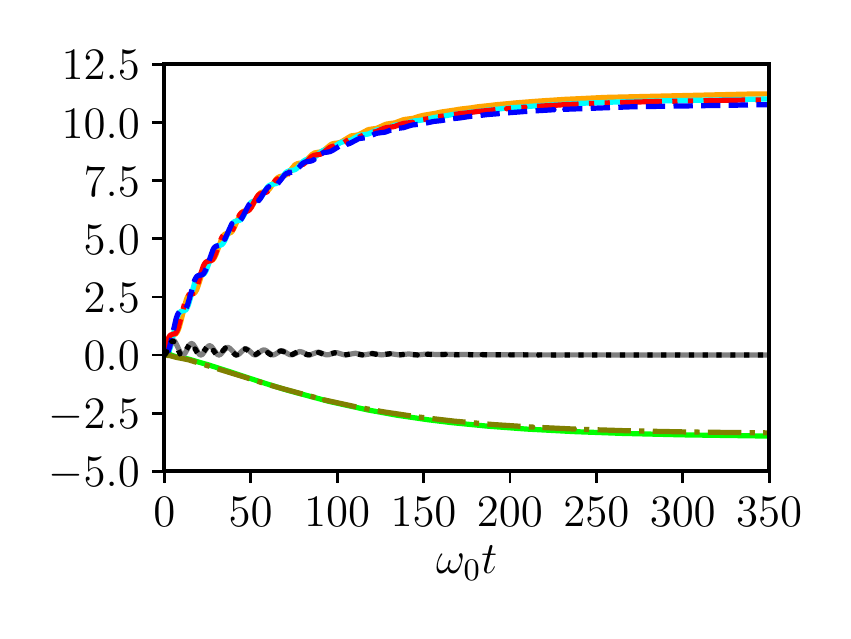}\put(30,73){}
\put(-20,75){CP-Redfield}
\end{overpic}
\\
\vspace{.5cm}
\begin{overpic}[width=0.49\linewidth]{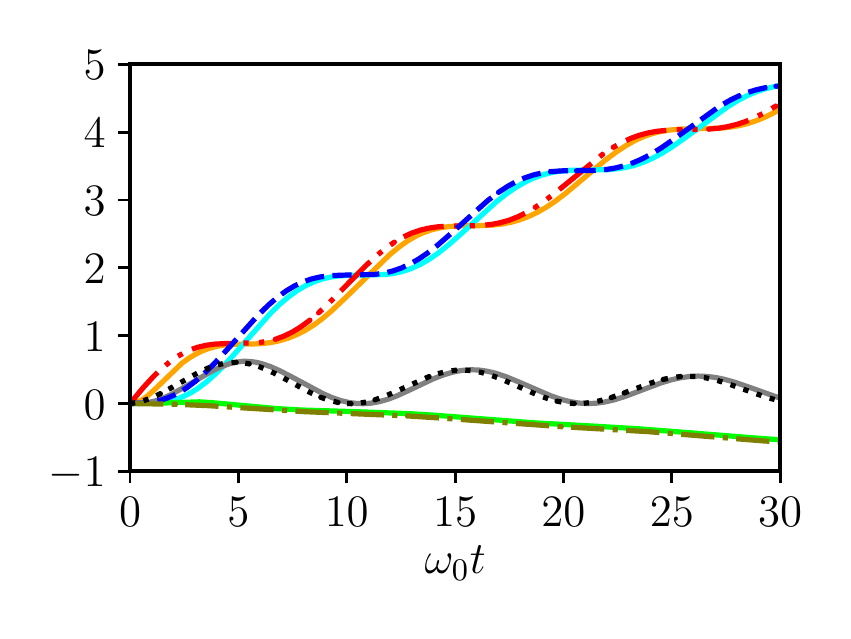}\put(30,73){(b)}\end{overpic}
\begin{overpic}[width=0.49\linewidth]{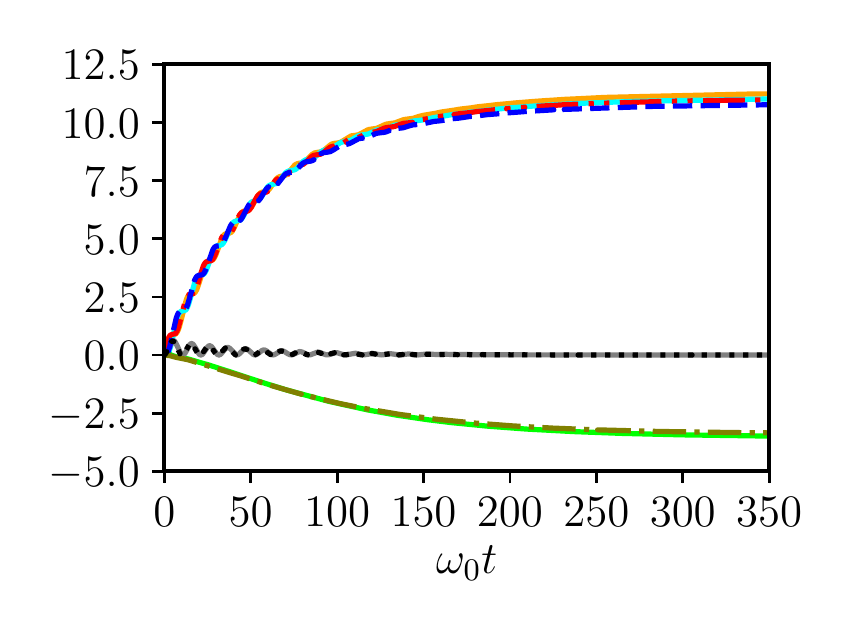}\put(30,73){}\put(-15,75){Redfield}
\end{overpic}
\caption{\label{fig:moms-redf}
(Color online) 
Comparison between second order moments evaluated using the  CP-Redfield (a)
and Redfield (b) with the ones predicted by the exact dynamics. 
As in the case of Fig.~\ref{fig:moms-glob-loc}
 continuous lines  represent
the quantities computed by solving the exact ${\cal S} + {\cal E}$ Hamiltonian
model~(\ref{hamiltonian-decomposition}) while dotted and dashed lines instead
refer to the approximated solutions.
%
Also each panel contains two plots corresponding each to shorter (left) and longer (right) time scales.
In all the plots we used    $\mathcal{N}(\omega_0)=10$ (corresponding to $1/\beta \approx10.5 \omega_0$),
$g=0.3 \omega_0$,  $\kappa(\omega_0)=0.04 \omega_0$,  $\omega_c=3 \omega_0$, $\alpha=1$ -- same as those used in Fig.~\ref{fig:moms-glob-loc}.
The value of $\Delta t$ used to define CP-Redfield is such that ${S}^{(\Delta t)}_{+-}=0.989~,$ which ensures the saturation of the inequality (\ref{eq:positivity-requirement}).
}
\end{figure*}

The convex combination (\ref{eq:convex-comb}) is not the only way of keeping the best from both the local and the global approximations.
Indeed, by making a step back, one can consider the coarse-grained Redfield equations~(\ref{eqs-Redf-sinc}) once that the pathology related to their non-positivity has been cured.
A detailed study of the performances of this approach is presented in Fig.~\ref{fig:moms-redf}. Here, for the same values of the parameters used in Fig.~\ref{fig:moms-glob-loc}, in panel (a) we exhibit the plots associated with the CP-Redfield equation obtained  by fixing ${S_{+ -}^{(\Delta t)}}$
in such a way to saturate the positivity bound~(\ref{eq:positivity-requirement}), i.e.
${S}^{(\Delta t)}_{+-}=0.989$.
As in the case of  panel (c)  of Fig.~\ref{fig:moms-redf}, we notice that
CP-Redfield is in a good agreement with the exact data both at long and short
time scales. As a check in panel (b) of Fig.~\ref{fig:moms-redf}
we also present the
(uncorrected)
Redfield equation obtained by setting in Eq.~(\ref{eqs-Redf-sinc}) $\Delta t=0$, corresponding to have ${S_{+ -}^{(\Delta t)}}=1$ which for the system parameters we choose
 gives a clear violation of the positivity bound~(\ref{eq:positivity-requirement}).
Interestingly enough, despite the fact that
the resulting equation does not guarantee complete positivity of the associated
evolution, we notice that also in this case one has an apparent good agreement with the exact results for all times. 
\begin{figure}
\begin{overpic}[width=0.98\linewidth]{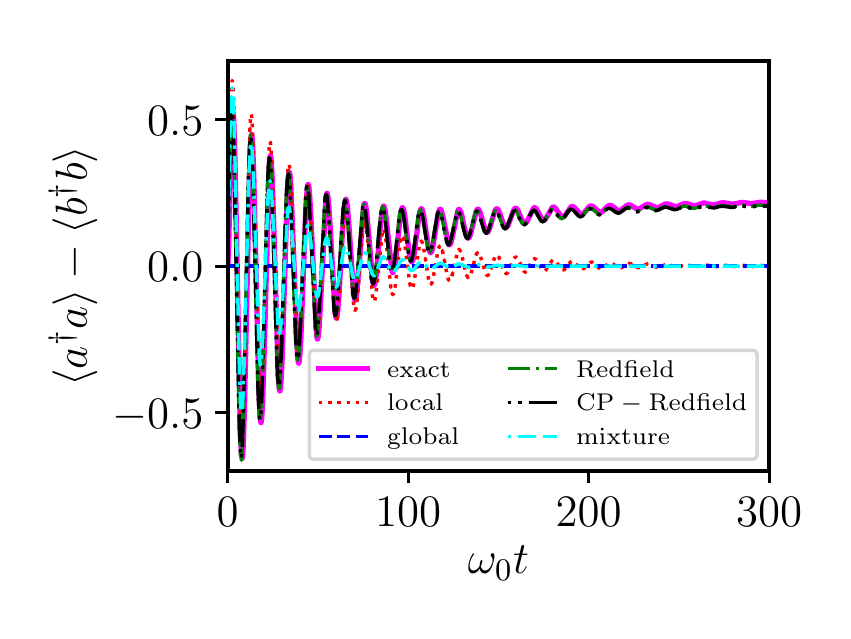}\end{overpic}
\caption{\label{fig:gap-A-B}
(Color online) 
Plot of the local excitation gap $\langle a^\dag a\rangle(t)-\langle b^\dag b\rangle(t)$ for the different approximation methods and for the exact dynamics.
Global ME (blue dashed line), Local ME (red dotted line), and 
convex Mixture approach (cyan dot-dashed-dashed line) predict an asymptotically
zero value for such quantities. On the contrary 
Redfield (green dot-dashed line) and 
CP-Redfield (black dot-dot-dashed line) give an asymptotic final value for such quantity
in agreement with the exact dynamics (magenta full and thicker line).
In all the plots we used    $\mathcal{N}(\omega_0)=10$ (corresponding to $1/\beta \approx10.5 \omega_0$),
$g=0.3 \omega_0$,  $\kappa(\omega_0)=0.04 \omega_0$,  $\omega_c=3 \omega_0$, $\alpha=1$ -- same as those used in Figs.~\ref{fig:moms-glob-loc}, \ref{fig:moms-redf}.
The value of $\Delta t$ used to define CP-Redfield is such that ${S}^{(\Delta t)}_{+-}=0.989~,$ which ensures the saturation of the inequality (\ref{eq:positivity-requirement}).}
\end{figure}
In particular both CP-Redfield and Redfield equations appear to be able to capture a non-weak coupling correction to the asymptotic value of $
2 {\rm Re} \langle \gamma_- \gamma_+^\dag \rangle (t)=
\langle a^\dag a\rangle(t)-\langle b^\dag b\rangle(t)~,$
an effect that is present in the exact model due to the fact that the
subsystem A remains slightly correlated with the bath degrees of freedom,  but which
is not present when adopting neither Global,  Local, or Mixed approximations (see Fig.~\ref{fig:gap-A-B}). 
An evidence of this can be
obtained by observing that  from Eq.~(\ref{eqs-Redf-sinc}) we have
\begin{eqnarray}
\label{eq:gapAB}
&&2 {\rm Re} \langle \gamma_- \gamma_+^\dag \rangle (\infty)=
\tfrac{S_{+ -}^{(\Delta t)}}{\omega_+ -\omega_-}
\\
&&\times 
\dashint_0^\infty d\epsilon ~\frac{\kappa(\epsilon)}{2 \pi}~ \left({\tfrac{\mathcal{N}(\epsilon)-\mathcal{N}(\omega_+) }{\epsilon- \omega_+} - \tfrac{\mathcal{N}(\epsilon)-\mathcal{N}(\omega_-) }{\epsilon- \omega_-}}\right)+O[\kappa(\omega_0)^2] \nonumber
\end{eqnarray}
which is exactly null for the Global ME ($S_{+ -}^{(\Delta t)}=0$), but which 
 is different from zero (and in  good agreement with the exact result) both
 for the uncorrected Redfield equation ($S_{+ -}^{(\Delta t)}=1$) and 
 CP-Redfield (${S}^{(\Delta t)}_{+-}=0.989$).

\begin{figure*}
\vspace{.5cm}
\begin{overpic}[width=.49\linewidth]{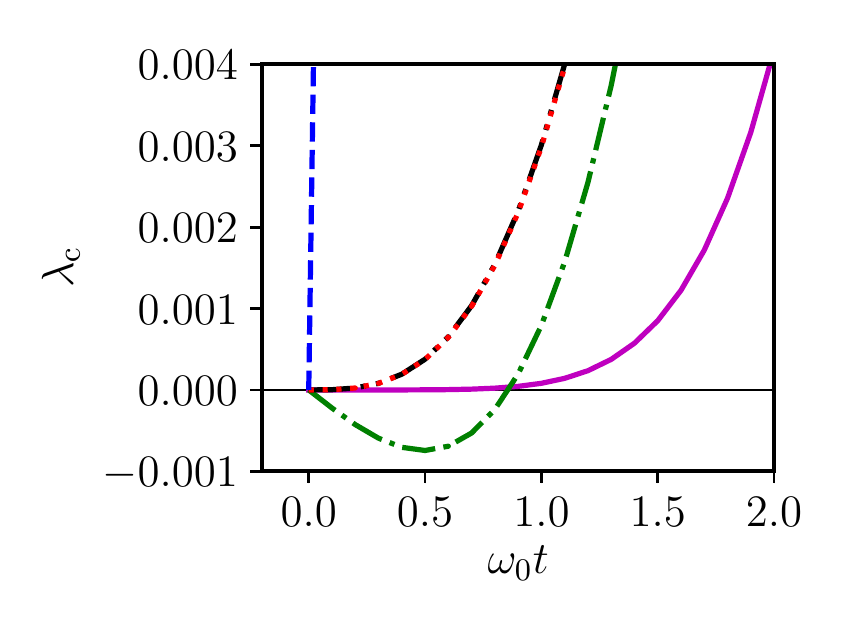}\put(5,75){(a)}\end{overpic}
\begin{overpic}[width=.49\linewidth]{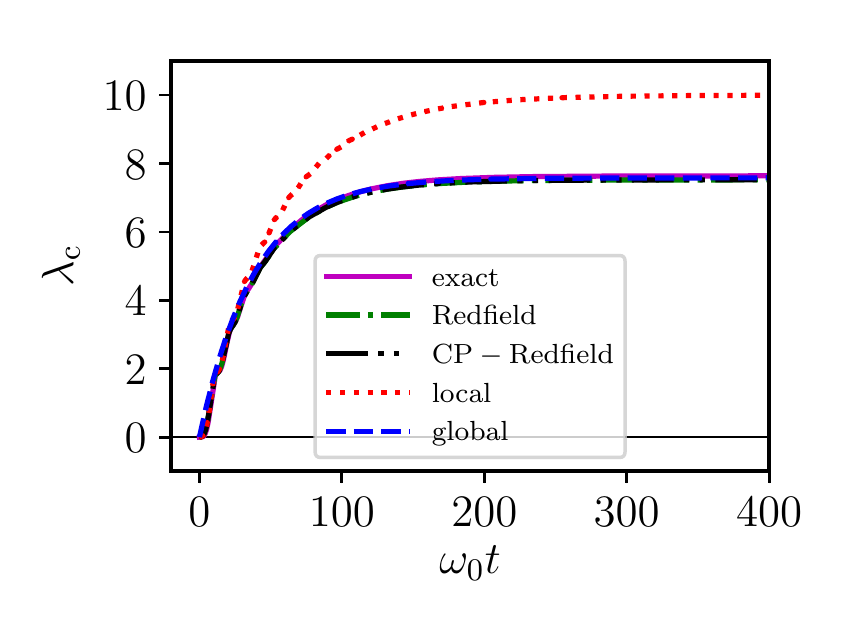}\put(5,75){(b)}\end{overpic}
\caption{\label{fig:avalc}
(Color online) 
Plots of the quantity $\lambda_{\rm c}(t)$ of Eq.~(\ref{eq:avalc}) for different approximation methods and using the exact result,
at shorter (a) and longer (b) time scales.
In all the plots we used    $\mathcal{N}(\omega_0)=10$ (corresponding to $1/\beta \approx10.5 \omega_0$),
$g=0.3 \omega_0$,  $\kappa(\omega_0)=0.04 \omega_0$,  $\omega_c=3 \omega_0$, $\alpha=1$ -- same as those used in Figs.~\ref{fig:moms-glob-loc}-\ref{fig:gap-A-B}. 
The value of $\Delta t$ used to define CP-Redfield is such that ${S}^{(\Delta t)}_{+-}=0.989~,$ which saturates the inequality (\ref{eq:positivity-requirement}).}
\end{figure*}
Despite the  apparent success of the uncorrected Redfield equation reported above, a clear signature of its non-positivity can still be spotted by looking 
at a special functional  of the second order moments of the model, i.e. the quantity \begin{equation}
\label{eq:avalc}
\lambda_c (t) :=   \frac{1}{2}{\rm min\{ ~eigenvalues} [ \Gamma_{\rm S}(t)+i \Xi_{\rm S} ]~\}~.
\end{equation}
In the above  definition 
 $\Gamma_{\rm S}(t)$ and $\Xi_{\rm S}$ are 
 respectively  the covariance matrix and the symplectic form of the two-mode system ${\cal S}$.
 Expressed in terms of the eigenoperators $\gamma_\pm$ their elements are given by 
 \begin{eqnarray} \label{COVXI} 
[{\Gamma}_{\rm S}(t)]_{ij}:= \left\langle \Big[ \pmb{\Gamma}_i -\langle \pmb{\Gamma}_i\rangle(t),\pmb{\Gamma}_j^\dag -\langle \pmb{\Gamma}^\dag_j\rangle(t)\Big]_+\right\rangle(t),
\end{eqnarray}
and 
 \begin{equation}\label{SIMXI} 
[\Xi_{\rm S}]_{ij} :=  -i \left\langle \Big[ \pmb{\Gamma}_i, \pmb{\Gamma}_j^\dag\Big]_- \right\rangle(t) =
-i 
\begin{pmatrix}
1 & 0 & 0 & 0\\
0 & -1 & 0 & 0\\
0 & 0 & 1 & 0\\
0 & 0 & 0 & -1
\end{pmatrix},
\end{equation}
 with $\pmb{\Gamma}_i$ being the $i$-th component of the operator vector $\pmb{\Gamma}:=(\gamma_+,\gamma_+^\dag,\gamma_-,\gamma_-^\dag)^T$. 
 In particular due to the choice of the input state we made in Eq.~(\ref{eq:in-cond-universe}),
 we get 
 \begin{widetext} 
 \begin{eqnarray}
\Gamma_{\rm S}(t) =
\begin{pmatrix}
2 \langle \gamma_+^\dag \gamma_+ \rangle (t)+1 & 0  & 2 \langle \gamma_- \gamma_+^\dag \rangle(t)^* & 0 \\
0 & 2 \langle \gamma_+^\dag \gamma_+ \rangle (t)+1 & 0 & 2 \langle \gamma_- \gamma_+^\dag \rangle(t)\\
2 \langle \gamma_- \gamma_+^\dag \rangle(t) & 0  & 2 \langle \gamma_-^\dag \gamma_- \rangle (t)+1 & 0\\
0  & 2 \langle \gamma_- \gamma_+^\dag \rangle(t)^* & 0 & 2 \langle \gamma_-^\dag \gamma_- \rangle (t)+1
\end{pmatrix}
~,
\end{eqnarray}
and hence
 \begin{eqnarray}
\frac{\Gamma_{\rm S}(t) + i \Xi_{\rm S}}{2}
 =
\begin{pmatrix}
 \langle \gamma_+^\dag \gamma_+ \rangle (t)+1 & 0  &  \langle \gamma_- \gamma_+^\dag \rangle(t)^* & 0 \\
0 &  \langle \gamma_+^\dag \gamma_+ \rangle (t) & 0 &  \langle \gamma_- \gamma_+^\dag \rangle(t)\\
 \langle \gamma_- \gamma_+^\dag \rangle(t) & 0  &  \langle \gamma_-^\dag \gamma_- \rangle (t)+1 & 0\\
0  &  \langle \gamma_- \gamma_+^\dag \rangle(t)^* & 0 &  \langle \gamma_-^\dag \gamma_- \rangle (t)
\end{pmatrix}
~,\\
\lambda_c(t)=\frac{1}{2}\{\langle \gamma_+^\dag \gamma_+ \rangle(t)+
\langle \gamma_-^\dag \gamma_- \rangle(t)-
\sqrt{[\langle \gamma_+^\dag \gamma_+ \rangle(t)-
\langle \gamma_-^\dag \gamma_- \rangle(t)]^2+4 |\langle \gamma_- \gamma_+^\dag  \rangle(t)|^2}
\}
~.
\end{eqnarray}
 \end{widetext} 
  When evaluated on a proper state of the system, the  
 Robertson-Schr\"{o}dinger uncertainty relation inequality~\cite{serafini2017quantum} 
 forces  the spectrum of the above matrix to be non-negative --
 see Appendix \ref{covariancesection} for details.
Accordingly 
 when  $\rho_{\rm S}(t)$ is positive semi-definite (i.e. it is a physical state) one must have $\lambda_c (t)\geq 0$.
The temporal evolution  of $\lambda_c (t)$ is reported in 
 Fig.~\ref{fig:avalc} for the various approximation methods and for the exact dynamics: one notice that while Global, Local, and CP-Redfield always complies with 
 the positivity requirement, the uncorrected 
 Redfield equation exhibit negative values of $\lambda_c(t)$ at short time scales.
Analytically, this can  
 be seen from the short time scale trend of $\lambda_c (t)~,$ which from 
 Eq.~(\ref{eqs-Redf-sinc}) can be determined as
\begin{eqnarray}
\label{eq:lambdaC-around0}
\lambda_c(\delta t)&\simeq&\left(\tfrac{\gamma_{--}^{(1)}+\gamma_{++}^{(1)}}{2} \right)
\Big[1-\\ \nonumber
&&\sqrt{1+
4
\big({S_{+ -}^{(\Delta t)}}^2-
\tfrac{\gamma_{++}^{(1)}\gamma_{--}^{(1)}}{|\gamma_{+-}^{(1)}|^2}
\big)
\tfrac{|\gamma_{+-}^{(1)}|^2}{(\gamma_{--}^{(1)}+\gamma_{++}^{(1)})^2}}~ \Big]~
\delta t\;,
\nonumber
\end{eqnarray}
which tightly gives $\lambda_c(\delta t)\geq  0$ if and only if the complete positivity constraint~(\ref{eq:positivity-requirement}) is fulfilled.
 Notice also that while  none of the approximated methods
 are able to follow the whole exact behaviour $\lambda_c(t)$, CP-Redfield and Global provide good agreement in the long time limit, while CP-Redfield and Local correctly 
predict $\dot{\lambda_c}(0)=0$. 

\begin{figure*}
\begin{overpic}[width=0.49\linewidth]{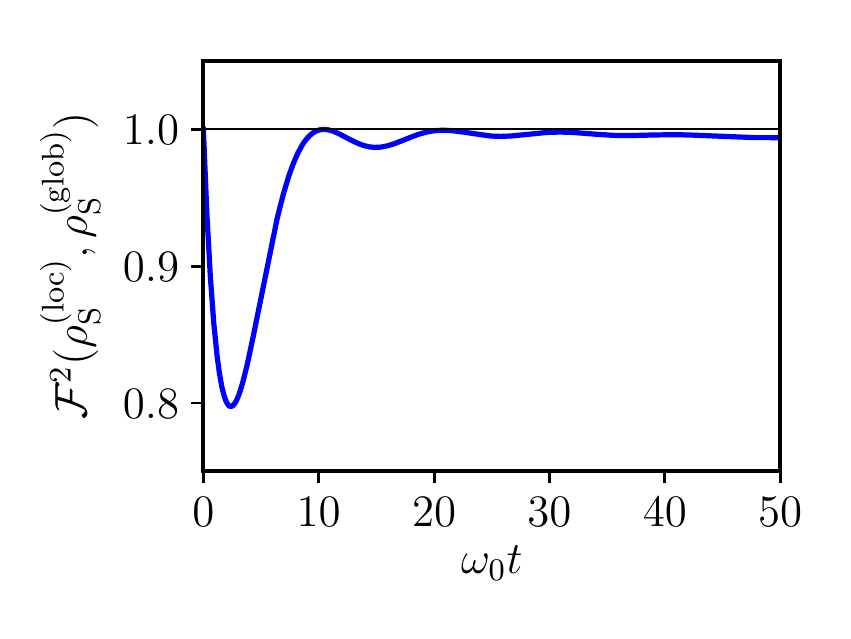}\put(30,73){(d)}\end{overpic}
\begin{overpic}[width=0.49\linewidth]{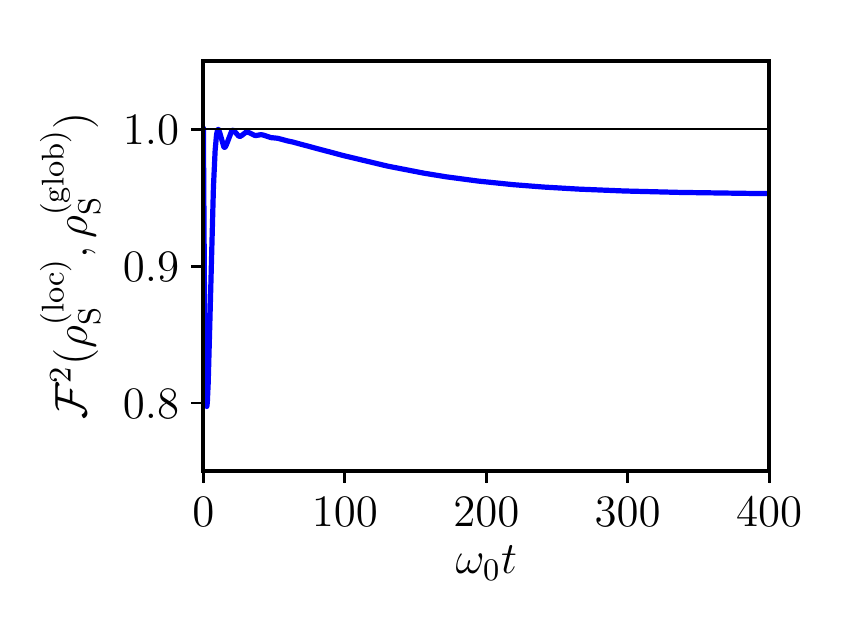}\put(30,73){}\end{overpic}

\caption{\label{fig:fidelity-loc-vs-glob}
Fidelity between the solutions $\rho^{(\rm loc)}_{\rm S}(t)$ and $\rho^{(\rm glob)}_{\rm S}(t)$ 
of the Local and Global MEs associated 
with the initial condition~(\ref{eq:in-cond-universe})
 at shorter (left) and longer (right) time scales.
In  the plots we used    $\mathcal{N}(\omega_0)=10$ (corresponding to $1/\beta \approx10.5 \omega_0$),
$g=0.3 \omega_0$,  $\kappa(\omega_0)=0.04 \omega_0$,  $\omega_c=3 \omega_0$, $\alpha=1$ -- same as those of Figs.~\ref{fig:moms-glob-loc}-\ref{fig:avalc}.}
\end{figure*}

\subsection{Fidelity Comparison}\label{FIDESEC} 
In this section we further discuss the difference between the various
approximation methods, as well as their relation with the exact solution,
evaluating the temporal evolution of the Uhlmann fidelity~\cite{nielsen-chuang-book} between the associated
density matrices of ${\cal S}$. 
We remind that given $\rho^{(1)}_{\rm S}$ and $\rho^{(2)}_{\rm S}$ two quantum states of the system their fidelity
is defined as the positive functional  
\begin{eqnarray}
\label{eq:fidelity-general}
\mathcal{F}(\rho^{(1)}_{\rm S}, \rho^{(2)}_{\rm S}):=\left\| \sqrt{\rho^{(1)}_{\rm S}} \sqrt{\rho^{(2)}_{\rm S}}\right\|_1\;,
\end{eqnarray}
with $\| \Theta \|_1 := \mbox{Tr}[ \sqrt{\Theta^\dag \Theta}]$ being the trace norm
of the operator $\Theta$. This quantity provides a {\it bona-fide} estimation of how close
the two density matrices are, getting its maximum value 1 when $\rho^{(1)}_{\rm S}=\rho^{(2)}_{\rm S}$,
and achieving zero value instead  when the support of
$\rho^{(1)}_{\rm S}$ and $\rho^{(2)}_{\rm S}$  are orthogonal, i.e. when they are perfectly distinguishable. 
In the case of two-mode Gaussian states~\cite{serafini2017quantum} with null first order moments, a relatively simple closed expression for $\mathcal{F}(\rho^{(1)}_{\rm S}, \rho^{(2)}_{\rm S})$ is known in terms of the covariance matrices of the two density matrices ~\cite{PhysRevA.86.022340, adesso2017loc-vs-glob, hofer2017markovian}.
Specifically, in the eigenmode representation, one has 
\begin{equation}
\mathcal{F}^2(\rho^{(1)}_{\rm S}, \rho^{(2)}_{\rm S})=\frac{1}{\sqrt{\rm b}+\sqrt{\rm c}-\sqrt{(\sqrt{\rm b}+\sqrt{\rm c})^2-{\rm a}
}}~, \label{OURVERSION} 
\end{equation}
with %
\begin{eqnarray}
\label{OURVERSION1} 
{\rm a}&:=&2^{-4} \det [\Gamma^{(1)}_{\rm S}+\Gamma^{(2)}_{\rm S}]~,
\\
\nonumber
{\rm b}&:=&2^{-4}\det[  \Xi_{\rm S}  \; \Gamma^{(1)}_{\rm S} \; \Xi_{\rm S}\;  \Gamma^{(2)}_{\rm S} -   \mathds{1}_4 ]
~,
\\
\nonumber
{\rm c} &:=& 2^{-4}\det[\Gamma^{(1)}_{\rm S}+i \Xi_{\rm S}]\det[\Gamma^{(2)}_{\rm S}+i \Xi_{\rm S}]
~,
\end{eqnarray}
where  $\Gamma^{(1)}_{\rm S}$, $\Gamma^{(2)}_{\rm S}$ being the covariance matrices of $\rho^{(1)}_{\rm S}$ and $\rho^{(2)}_{\rm S}$ defined in 
 (\ref{COVXI}), and with 
$\Xi_{\rm S}$ the
symplectic form given in  Eq.~(\ref{SIMXI}) -- see final part of Appendix~\ref{covariancesection} for details.
In what follows we shall make extensive use of the
identity~(\ref{OURVERSION})
 thanks to the fact that 
for the input
state~(\ref{eq:in-cond-universe}) we are considering in our analysis, 
the density matrix  of ${\cal S}$ remains Gaussian at all times when evolved 
under 
Global, Local, CP-Redfield ME, as well as under the exact integration of the full
${\cal S} +{\cal E}$ Hamiltonian model. The same property unfortunately does not hold for the convex mixture~(\ref{eq:convex-comb}) which is explicitly non-Gaussian (indeed it is a convex combination of Gaussian states). In this case
hence the result of ~\cite{PhysRevA.86.022340} can not be directly applied to
compute $\mathcal{F}\left(\rho^{(\rm mix)}_{\rm S}(t) ,\rho^{(\rm exact)}_{\rm S}(t)\right)$.
Still  the {concavity} property \cite{nielsen-chuang-book} of the $\mathcal{F}$ 
can be invoked to compute the following  lower bound  
\begin{eqnarray}
&&\mathcal{F}\left(\rho^{(\rm mix)}_{\rm S}(t) ,\rho^{(\rm exact)}_{\rm S}(t)\right)
\geq
e^{- \mathcal{G} t}  \mathcal{F}\left(\rho^{(\rm loc)}_{\rm S}(t) ,\rho^{(\rm exact)}_{\rm S}(t)\right)
\nonumber \\
&&\qquad \qquad\qquad +
(1-e^{- \mathcal{G} t}) \mathcal{F}\left(\rho^{(\rm glob)}_{\rm S}(t) ,\rho^{(\rm exact)}_{\rm S}(t)\right),
\label{eq:fidelity-bound-convex-mixture}
\end{eqnarray}  
with  the  right-hand-side  being provided by Gaussian terms.
Finally 
the non-positivity of the (uncorrected) Redfield equation also gives rise to problems in the evaluation of the associated fidelity (as a matter of fact, in this case 
the quantity $\mathcal{F}\left(\rho^{(\rm red)}_{\rm S}(t) ,\rho^{(\rm exact)}_{\rm S}(t)\right)$ is simply ill defined). 
Aware of this fundamental limitation, but also of the  fact that the departure from the
positivity condition  of the solution $\rho^{(\rm red)}_{\rm S}(t)$ of the Redfield equation 
is small, in our analysis we decided to present the real part of $\mathcal{F}^2\left(\rho^{(\rm red)}_{\rm S}(t) ,\rho^{(\rm exact)}_{\rm S}(t)\right)$.

\begin{figure*}
\begin{overpic}[width=0.49\linewidth]{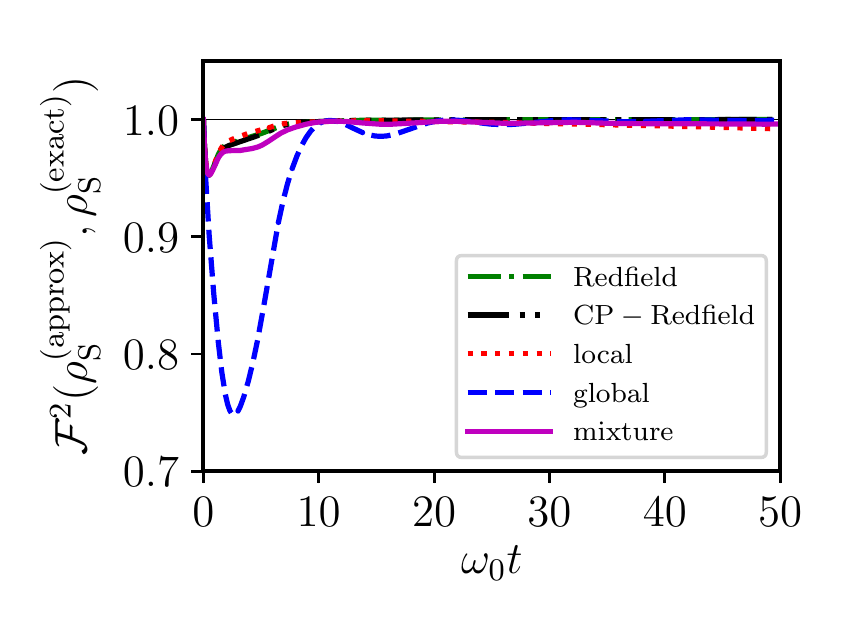}\put(5,75){(a)}\end{overpic}
\begin{overpic}[width=0.49\linewidth]{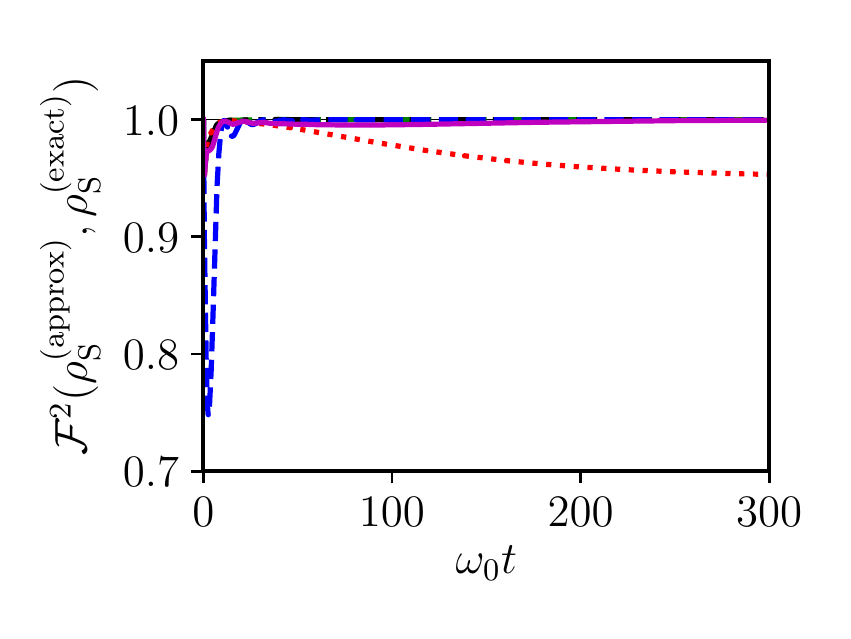}\put(5,75){(b)}\end{overpic}
\begin{overpic}[width=0.49\linewidth]{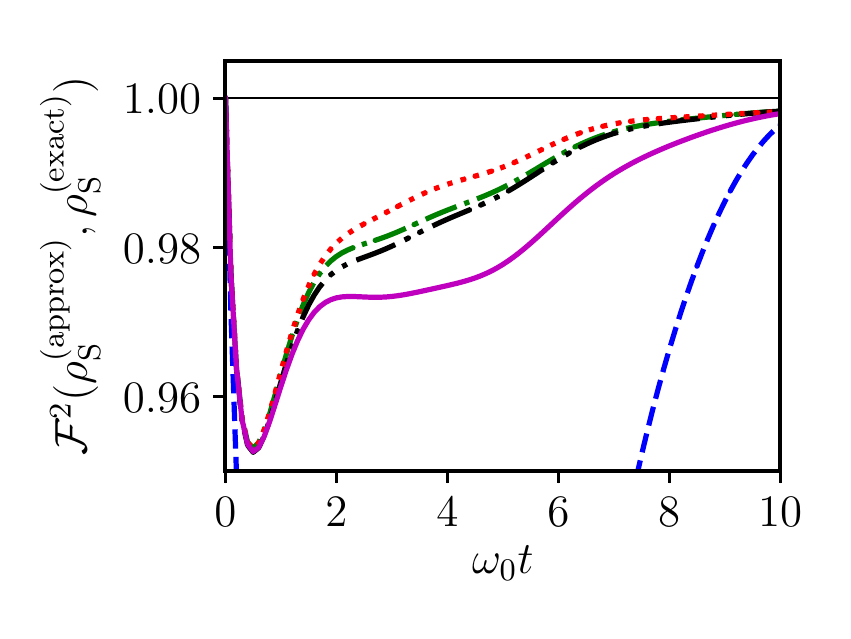}\put(5,75){(c)}\end{overpic}
\begin{overpic}[width=0.49\linewidth]{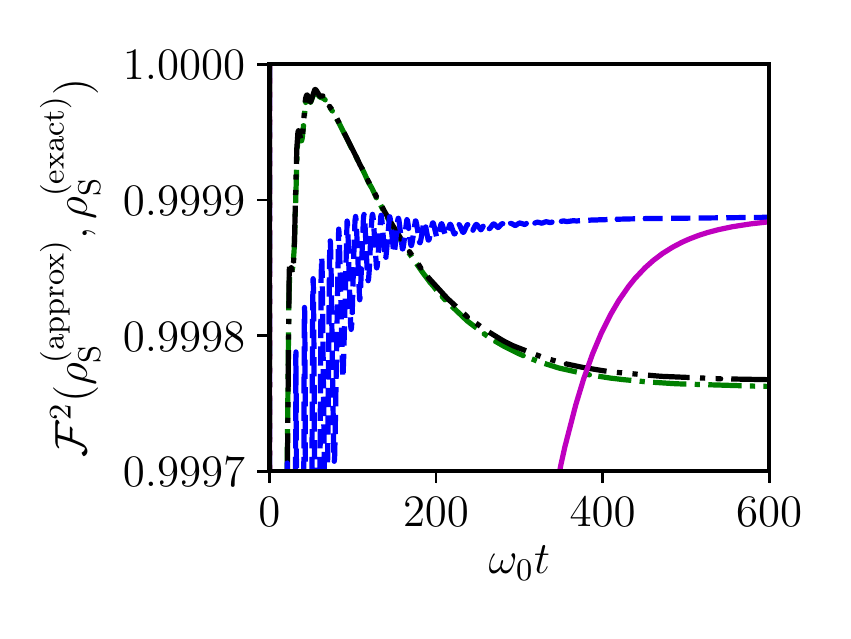}\put(5,75){(d)}\end{overpic}
\caption{\label{fig:fidelity}
(Color online) 
Fidelity between approximated system states and the exact system state. 
Different curves refer to the kind of approximation (see the legend): 
Redfield, green dot-dashed line (using ${\rm Re}(\mathcal{F}^2)$);
CP-Redfield, black dot-dot-dashed line;
local, red dotted line;
global, blue dashed line;
convex mixture
of Eq.~(\ref{eq:convex-comb}) with $\mathcal{G}=0.4 \kappa(\omega_0)$, 
magenta full line (using the lower bound given in the right-hand-side of Eq.~(\ref{eq:fidelity-bound-convex-mixture})).
The four panels differ just for the axes scales.
In  the plots we used    $\mathcal{N}(\omega_0)=10$ (corresponding to $1/\beta \approx10.5 \omega_0$),
$g=0.3 \omega_0$,  $\kappa(\omega_0)=0.04 \omega_0$,  $\omega_c=3 \omega_0$, $\alpha=1$ -- same as those of Figs.~\ref{fig:moms-glob-loc}-\ref{fig:fidelity-loc-vs-glob}.}
\end{figure*}

\begin{figure*}
\begin{overpic}[width=0.49\linewidth]{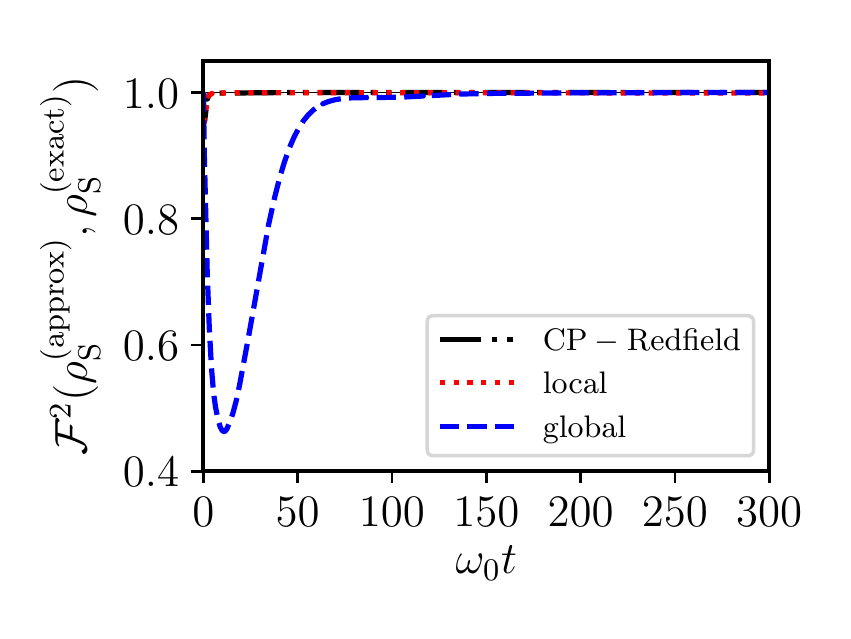}\put(5,75){(a)}\end{overpic}
\begin{overpic}[width=0.49\linewidth]{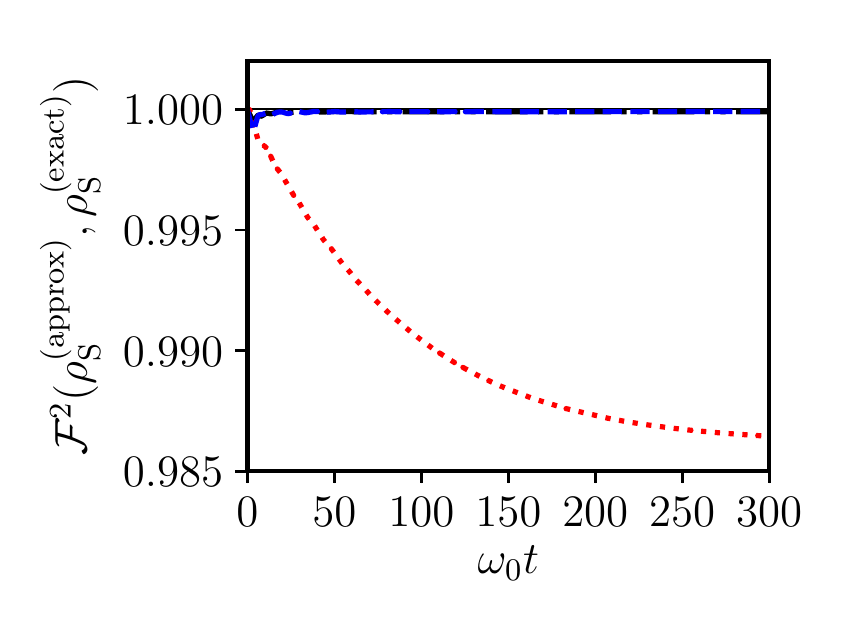}\put(5,75){(b)}\end{overpic}
\caption{\label{fig:fidelity-other-parameters}
(Color online) 
Fidelity between approximated system states and the exact system state for 
(a)~$\mathcal{N}(\omega_0)=10$, $g=0.04 \omega_0$
(weaker internal coupling) 
and 
(b)~$\mathcal{N}(\omega_0)=0.01$ (low temperature regime),  $g=0.3 \omega_0$.
As explained in the legend the black dot-dot-dashed line refer to the 
CP-Redfield solution (${S}^{(\Delta t)}_{+-}=0.9998$ in (a) and ${S}^{(\Delta t)}_{+-}=0.4813$ in (b)); the red dotted line to the Local ME solutions, and finally 
the blue dashed line to the 
Global ME solution. In all the plots we assumed $\alpha=1$ (Ohmic  spectral density regime) and kept $\kappa(\omega_0)=0.04 \omega_0$,  $\omega_c=3 \omega_0$. 
Notice finally that in (b) the Fidelity is generally higher (see the different ordinate scales in (a) and (b)). This is due to the choice of the ground-state (\ref{eq:in-cond-system}) as initial state, which implies that at low temperature such initial condition is just weakly modified.}
\end{figure*}

To begin, in Fig.~\ref{fig:fidelity-loc-vs-glob} we present the value of
 $\mathcal{F}^2(\rho^{(\rm loc)}_{\rm S}(t), \rho^{(\rm glob)}_{\rm S}(t))$: as clear from the plot, this quantity 
is sensibly different from 1  at short and at long time scales (confirming the observation
of the previous section) while it is $\sim 1$ at intermediate time scales.
In Fig.~\ref{fig:fidelity} instead we proceed with the comparison of the
approximate solutions with the exact one. The reported plots  
confirm that the convex combination of the local and global solutions (\ref{eq:convex-comb}) is an effective ansatz to approximate the system evolution, giving a (lower) bound for the fidelity computed as in Eq.~(\ref{eq:fidelity-bound-convex-mixture})
that is close to 1 both at short and at long time scales.
On the same footing we find the CP-Redfield equation which still remaining positive brings all the main qualities of the (full) Redfield ME.
For completeness,  in Fig.~\ref{fig:fidelity-other-parameters} we report two situations in which the Global ME  and the local ME work extremely bad respectively. 
In Panel (a) we consider weaker internal coupling $g$ such that the local ME gives a satisfying result for the whole dynamics while the inadequacy of the Global ME during the transient is accentuated;
In Panel (b) we decrease instead the temperature accentuating the inadequacy of the local ME in the steady prediction.
In both the Panels we report the curve corresponding to the CP-Redfield approximation. The last 
follows either the local or the global curve depending on which one performs better in the two instances.

\section{Conclusions}
\label{sec:conclusions}

In the study of multipartite Markovian open quatum systems it has been widely discussed in literature whether the \textit{local} dissipator or the \textit{global} dissipator is more adapt to effectively  reproduce the system dynamics \cite{hofer2017markovian,rivas2010markovian,adesso2017loc-vs-glob,cattaneo2019psa}.
Here we have treated a case where the system is composed of two interacting harmonic oscillators A and B, with only A interacting with a thermal bath - collection of other harmonic oscillators - and we have analyzed the equilibration process of the system initially in the ground state with the finite bath temperature.
We have shown that the  ``completely positive Redfield'' equation --- i.e. the cured version of the Redfield equation by means of coarse-grain averaging \cite{farina2019psa} --- and an appropriate time-dependent convex mixture of the local and global solutions~(\ref{eq:convex-comb}) give rise to the most accurate semigroup approximations of the exact system dynamics, both during the time transient and for the steady state properties, going beyond the {pure} local and global approximations.
The convex mixture of the local and global channels has been introduced phenomenologically for allowing at the same time coherent local energy exchange at short time scales between A and B and the steady state  expected from the thermodynamics at long time scales, i.e. the global Gibbsian state. 
Future developments on this route may concern the search of a microscopic derivation of this (non-Markovian) quantum channel.

D. F and V. G. acknowledge support by MIUR via PRIN 2017 (Progetto di Ricerca di Interesse Nazionale): project QUSHIP (2017SRNBRK).

\pagebreak
\newpage
\appendix

\section{Derivation of the coarse-grained Redfield and Local ME}
\label{appendix:approximations}
In this section we provide details about the derivations of the coarse-grained Redfield (\ref{eq:redfield-coarse-grain-state}) and Local (\ref{eq:loc-me-Sch-pict}) MEs.
For (\ref{eq:redfield-coarse-grain-state}) we make use of Refs.~\cite{hofer2017markovian, breuer2002theory} and of the method to correct the non-positivity of the Redfield equation given in Ref.~\cite{farina2019psa}, while for (\ref{eq:loc-me-Sch-pict}) we follow the approach of Ref.~\cite{hofer2017markovian}. 

Expressed in interaction picture the evolution of the joint state of ${\cal S} + {\cal E}$ induced by the
 Hamiltonian~(\ref{hamiltonian-decomposition})  is given by the 
 Liouville-von Neumann equation
%
\begin{eqnarray}
\dot{\tilde{\rho}}_{\rm SE}(t)&=&-i \Big[  \tilde{H}_{1}(t), \tilde{\rho}_{\rm SE}(t)\Big]_-~,
\label{eq:meSEexact}
\end{eqnarray} 
where given $U_{\rm 0}(t):= e^{i (H_{\rm S}+H_{\rm E}) t}$ we have 
\begin{eqnarray} 
\tilde{H}_{1}(t)&:=&U_{\rm 0} (t) H_{1}  U_{\rm 0}^\dagger (t) =a^\dagger(t) C(t)+ h.c.
\end{eqnarray}
%
with $a^\dagger(t):= e^{i H_{\rm S,g}t} a^\dagger e^{-i H_{\rm S,g}t}$ and 
$C(t):= \sum_k \gamma_k c_k e^{-i(\omega_k - \omega_0)t}$.
%
%
Tracing out the environment degrees of freedom, Eq.~(\ref{eq:meSEexact}) can be written as
\begin{eqnarray}\label{eq:meSexact}
\dot{\tilde{\rho_{\rm S}}}(t)&=&-i {\rm Tr}_{\rm E}\Big[ \tilde{H}_{\rm 1}(t), \tilde{\rho}_{\rm SE}(0)\Big]_-
\\&&-\int_0^t {\rm Tr}_{\rm E}\Big[ \tilde{H}_{\rm 1}(t),\Big[ \tilde{H}_{\rm 1}(t'), \tilde{\rho}_{\rm SE}(t')\Big]_-\Big]_-dt' 
~. \nonumber 
\end{eqnarray} 
We assume now weak system-environment coupling such that the environment stays in its own Gibbs state~(\ref{iniENV})  (invariant in interaction picture) for all the system dynamics and the SE state can be approximated by the tensor product 
\begin{equation}
\label{eq:tensor-product-SE}
\tilde{\rho}_{\rm SE}(t)\simeq\tilde{\rho}_{\rm S}(t) \otimes {\rho}_{\rm E}(0).
\end{equation}
%
%
Equation~(\ref{eq:tensor-product-SE}) means that the environment, being a macroscopic object, can be considered insensitive to the interaction with the system (Born approximation \cite{breuer2002theory}). On the contrary the system state is affected by the coupling with the environment. Being the first moments null over a thermal state, the first commutator in (\ref{eq:meSexact}) is zero and by inserting the tensor product (\ref{eq:tensor-product-SE}) such equation becomes
\begin{eqnarray}
&& \label{eq:meSBornc1c2}
\dot{\tilde{\rho}}_{\rm S}(t)\simeq
\\
&&
\int_0^t dt'  c^{(1)}(t-t')
\left(a^\dagger (t') \tilde{\rho}_{\rm S}(t') a(t)-
a(t) a^\dagger(t') \tilde{\rho}_{\rm S}(t')\right)+
\nonumber\\ 
&&
c^{(2)}(t-t') \left(a(t') \tilde{\rho}_{\rm S}(t') a^\dagger(t) -
a^\dagger(t) a(t') \tilde{\rho}_{\rm S}(t')\right) + h.c.\,, \nonumber
\end{eqnarray}
where $c^{(1)}(\tau)$ and $c^{(2)}(\tau)$ are bath correlation functions defined as
\begin{eqnarray}
c^{(1)}(\tau)&:=&\left<C^\dagger(\tau) C \right>=\sum_k \gamma_k^2 \mathcal{N}(\omega_k) e^{i (\omega_k-\omega_0)\tau}
~, \label{eq:c1c2}
\\
c^{(2)}(\tau)&:=&\left<C(\tau) C^\dagger \right>=\sum_k \gamma_k^2 [1+\mathcal{N}(\omega_k)] e^{-i (\omega_k-\omega_0)\tau}~.
\nonumber
\end{eqnarray}
Next step is the Markovian assumption $\tau_{\rm E}  \ll \delta t$,  where $\delta t$ is the typical time scale of the state in interaction picture and $\tau_{\rm E}$ is the bath memory time scale, i.e. the characteristic width of the bath correlation functions (\ref{eq:c1c2}).
Such time scale separation allows to replace in Eq.~(\ref{eq:meSBornc1c2}) the upper integration bound with $+\infty$ and to neglect the $\tau:= t-t'$ dependence of the state $\tilde{\rho}_{\rm S}$, 
leading to the Redfield equation (interaction picture):
\begin{eqnarray}\label{eq:redfield-int-pict}
&&
\dot{\tilde{\rho}}_{\rm S}(t)\simeq \int_0^\infty d\tau  \Big[ c^{(1)}(\tau)
\\&& \qquad \times
\left(a^\dagger (t-\tau) \tilde{\rho}_{\rm S}(t) a(t)-
a(t) a^\dagger(t-\tau) \tilde{\rho}_{\rm S}(t)\right)
\nonumber\\&&+
c^{(2)}(\tau) \left(a(t-\tau) \tilde{\rho}_{\rm S}(t) a^\dagger(t) -
a^\dagger(t) a(t-\tau) \tilde{\rho}_{\rm S}(t)\right) \Big]+ h.c.
\nonumber 
\end{eqnarray} 
As described in Ref.~\cite{hofer2017markovian},
if the bath correlation functions are narrow enough with respect to the internal coupling time scale, i.e. $g \tau_{\rm E}\ll 1,
$
in Eq.~(\ref{eq:redfield-int-pict})
one can approximate $a(t-\tau)\approx a(t)~$ 
obtaining the interaction picture version of the Local ME (\ref{eq:loc-me-Sch-pict}), which is in Lindblad form without the need of any secular approximation. Alternatively, passing to the eigenmode basis of Eq.~(\ref{eq:eigenmodes}), 
Eq.~(\ref{eq:redfield-int-pict}) can be equivalently written as  
\begin{eqnarray}
\nonumber
\dot{\tilde{\rho}}_{\rm S}(t)&=&
\frac{1}{2} \sum_{\sigma, \sigma'}  \Big[ 
\Omega^{\rm(1)}_{\sigma} 
e^{i (\sigma-\sigma')g t} 
\left(\gamma_{\sigma}^\dagger \tilde{\rho}_{\rm S}(t) \gamma_{\sigma'}-
\gamma_{\sigma'} \gamma_{\sigma}^\dagger \tilde{\rho}_{\rm S}(t)\right) 
\\ \nonumber &&+
\Omega^{\rm(2)}_{\sigma'}
e^{i (\sigma-\sigma')g t}
 \left(
 \gamma_{\sigma'} \tilde{\rho}_{\rm S}(t) \gamma_{\sigma}^\dagger -
\gamma_{\sigma}^\dagger \gamma_{\sigma'} \tilde{\rho}_{\rm S}(t)\right) \Big]
+
 h.c.\nonumber\\
\label{eq:redfield-global-basis}
\end{eqnarray} 
where
\begin{eqnarray} 
\Omega^{\rm(1)}_{\sigma}&:=&\int_0^\infty d\tau  c^{(1)}(\tau) e^{-i \sigma g \tau}\;, \\ 
\Omega^{\rm(2)}_{\sigma'}&:=&
\int_0^\infty d\tau c^{(2)}(\tau) e^{i \sigma' g \tau}\;.
\end{eqnarray} 
Next step is to
perform a coarse-grain average on Eq.~(\ref{eq:redfield-global-basis}) over a time interval $\Delta t\ll \delta t$, 
which amounts in applying the following substitution 
\begin{eqnarray}
e^{i (\sigma-\sigma')g t}
&\longrightarrow&
 \frac{1}{\Delta t} \int_{t-\Delta t/2}^{t+\Delta t/2} ds ~e^{i (\sigma-\sigma')g s}\nonumber\\&=&e^{i (\sigma-\sigma')g t} 
 {\rm sinc}\left(\frac{(\sigma-\sigma')g \Delta t}{2}\right),
\end{eqnarray}
without affecting the system state in interaction picture.
Equation~(\ref{eq:redfield-coarse-grain-state}) is eventually obtained by passing to the Schr\"{o}dinger picture. 
Indeed the Lamb-shift and the dissipator coefficients of Eqs.~(\ref{eq:gamma-eta-sigma-sigma'}) and 
(\ref{eq:gamma-eta-sigma-sigma'1}) are related to the quantities $\Omega^{(i)}_{\sigma}$ as 
\begin{eqnarray}
\gamma^{(i)}_{\sigma \sigma'}=\frac{1}{2} 
(\Omega^{(i)}_{\sigma}+{\Omega^{(i)}_{\sigma'}}^*)~,
\end{eqnarray}
\begin{eqnarray}
\eta^{(i)}_{\sigma \sigma'}=\frac{1}{4i} 
(\Omega^{(i)}_{\sigma}-{\Omega^{(i)}_{\sigma'}}^*)~.
\end{eqnarray}
%
%
%

\section{Completely positive map requirement for the coarse-grained Redfield equation}\label{CPSEC} 
To discuss the complete positivity condition for the coarse-grained Redfield equation let us observe that 
its  dissipator is given by  
the last two
lines in the right-hand-side of  Eq.~(\ref{eq:redfield-coarse-grain-state}). 
Following  Ref.~\cite{farina2019psa} we 
write them as
\begin{eqnarray}
\sum_{i, \sigma, i', \sigma'} 
{\gamma}_{i' \sigma' , i \sigma }
\left(\mathcal{A}_{i', \sigma'}^\dagger \rho_{\rm S}(t) \mathcal{A}_{i, \sigma}
- \frac{1}{2} \Big[ \mathcal{A}_{i, \sigma}\mathcal{A}_{i', \sigma'}^\dagger\,,\,\rho_{\rm S}(t) \Big]_+ \right),
\nonumber
\end{eqnarray}
with $\mathcal{A}_{1, \sigma}=\gamma_\sigma$, $
\mathcal{A}_{2, \sigma}=\gamma_\sigma^\dagger$,
and  ${\gamma}_{i' \sigma' , i \sigma }$ being the  elements of the  
$4 \times 4$ hermitian matrix 
\begin{equation}
{\gamma}_{I , J } 
=
 \begin{pmatrix} 
  \gamma^{\rm(1)}_{+ +}&\gamma^{\rm(1)}_{+ -}S_{+ -}^{(\Delta t)}&0&0\\
  \gamma^{\rm(1)}_{- +}S_{+ -}^{(\Delta t)}&\gamma^{\rm(1)}_{- -}&0&0\\
  0&0&\gamma^{\rm(2)}_{+ +}&\gamma^{\rm(2)}_{+ -}S_{+ -}^{(\Delta t)}\\
  0&0&\gamma^{\rm(2)}_{- +}S_{+ -}^{(\Delta t)}&\gamma^{\rm(2)}_{- -}\\  
\end{pmatrix}.
\label{eq:dissipation-matrix-psa}
\end{equation}
Complete positivity of  the evolution described by Eq.~(\ref{eq:redfield-coarse-grain-state}) can now be guaranteed by the imposing positiveness of the spectrum 
of  (\ref{eq:dissipation-matrix-psa}), a condition which  by explicit diagonalization  leads to 
Eq.~(\ref{eq:positivity-requirement}).

\section{Covariance matrices} \label{covariancesection}

Expressed in terms of the  system canonical coordinates 
 \begin{eqnarray}
 \label{eq:xA...pB}
 x_{\rm A}:=(a+a^\dag)/\sqrt{2}\;, &\qquad& p_{\rm A}:=(a-a^\dag)/(\sqrt{2} i)\;, \nonumber\\ 
x_{\rm B}:=(b+b^\dag)/\sqrt{2}\;, &\qquad& p_{\rm B}:=(b-b^\dag)/(\sqrt{2} i)\;,
\end{eqnarray} 
the covariance matrix $\Sigma_{\rm S}$  associated with the quantum state $\rho_{\rm S}$ of the two-mode system ${\cal S}$ is defined as the $4\times 4$ real hermitian  matrix
 \begin{equation} \label{COVAB} 
[\Sigma_{\rm S}]_{\alpha \beta}:= \left\langle \Big[ \pmb{r}_{{\rm S},\alpha} - \langle \pmb{r}_{{\rm S},\alpha} \rangle , 
\pmb{r}_{{\rm S},\beta}  - \langle \pmb{r}_{{\rm S},\beta}  \rangle\Big]_+
\right\rangle   \;, 
\end{equation}
 where  as usual we adopt the shorthand notation $\langle \cdots \rangle : = \mbox{Tr} [\cdots  \rho_{\rm S}]$,  and where $\pmb{r}_{{\rm S},\alpha}$ is the $\alpha$-th component of the operator vector 
 $\pmb{r}_{\rm S}:=(x_{\rm A}, p_{\rm A}, x_{\rm B}, p_{\rm B})^T$.
 In this notation 
 the symplectic form of the system is defined by the matrix   $\Omega_{\rm S}$ of elements
\begin{eqnarray} \label{SIMP2} 
\Omega_{\rm S}:=
\begin{pmatrix}
0 & 1 & 0 & 0\\
-1 & 0 & 0 & 0\\
0 & 0 & 0 & 1\\
0 & 0 & -1 & 0
\end{pmatrix}
~,
\end{eqnarray}
which embodies the 
canonical  commutation rules of the model via the identity 
$\langle [\pmb{r}_{{\rm S},\alpha},\pmb{r}_{{\rm S},\beta}]_- \rangle=i [\Omega_{\rm S}]_{\alpha \beta}$.
From the Robertson-Schr\"{o}dinger uncertainty relations \cite{serafini2017quantum} it hence follows, 
that for all choice of  $\rho_{\rm S}$ we must have that the matrix 
$\Sigma_{\rm S}+i \Omega_{\rm S}$ is non-negative or equivalently that the following inequality must hold
\begin{equation}
\label{eq:avalcnew}
{\rm min\{ ~eigenvalues} [ \Sigma_{\rm S}+i \Omega_{\rm S} ]~\}\geq 0\;.
\end{equation}
Equation~(\ref{eq:avalcnew}) is at the origin 
 of the study we presented in Fig.~\ref{fig:avalc}.
We notice indeed that introducing
 the unitary matrix 
 \begin{eqnarray}
 \mathcal{V} := \frac{1}{2}
\begin{pmatrix}
1 & 1 & 1 & 1\\
-i & i & -i & i\\
1 & 1 & - 1 & -1\\
-i & i & i & -i
\end{pmatrix}\;,
\end{eqnarray}
from  Eq.~(\ref{eq:eigenmodes})  the following identity holds, 
 \begin{eqnarray} 
 \pmb{r}_{\rm S} &=& {\cal V}  \pmb{\Gamma}_{\rm S}\;,  
 \end{eqnarray} 
 with $\pmb{\Gamma}_{\rm S}$ the operator vector introduced in Eq.~(\ref{COVXI}),
 which in turns implies
 \begin{eqnarray} 
   \Sigma_{\rm S} = {\cal V}  \Gamma_{\rm S} {\cal V}^\dag \;, \qquad 
    \Omega_{\rm S} = {\cal V}  \Xi_{\rm S} {\cal V}^\dag \;, \label{UNITARY21} 
    \end{eqnarray} 
    with $\Xi_{\rm S} $ as in Eq.~(\ref{SIMXI}). 
Accordingly, we get 
\begin{eqnarray} 
 \Sigma_{\rm S}+i \Omega_{\rm S} = {\cal V} \left( \Gamma_{\rm S}+i \Xi_{\rm S}\right) {\cal V}^\dag\;, 
\end{eqnarray} 
which finally allows us to translate Eq.~(\ref{eq:avalcnew}) into the positivity condition for the quantity
 $\lambda_c(t)$ introduced in Eq~(\ref{eq:avalc}).

Notice finally that the unitary relations (\ref{UNITARY21}) are also at the origin of 
Eqs.~(\ref{OURVERSION}) and (\ref{OURVERSION1}) which we derived from~\cite{PhysRevA.86.022340,adesso2017loc-vs-glob, hofer2017markovian}
via the
 identities
\begin{eqnarray} 
\det [\Gamma^{(1)}_{\rm S}+\Gamma^{(2)}_{\rm S}] &=& \det [\Sigma^{(1)}_{\rm S}+\Sigma^{(2)}_{\rm S}] \;, \nonumber \\ \nonumber 
\det[  \Xi_{\rm S}  \; \Gamma^{(1)}_{\rm S} \; \Xi_{\rm S}\;  \Gamma^{(2)}_{\rm S} -   \mathds{1}_4 ]&=&
\det[  \Omega_{\rm S}  \; \Sigma^{(1)}_{\rm S} \; \Omega_{\rm S}\;  \Sigma^{(2)}_{\rm S} -   \mathds{1}_4 ]\;, \nonumber \\
\det[\Gamma^{(j)}_{\rm S}+i \Xi_{\rm S}]&=&
\det[\Sigma^{(j)}_{\rm S}+i \Omega_{\rm S}]\;, 
\end{eqnarray} 
where for $j=1,2$,  $\Gamma^{(j)}_{\rm S}$ and $\Sigma^{(j)}_{\rm S}$ represent the covariance matrices
(\ref{COVXI}) and (\ref{COVAB}) of the matrices $\rho_{\rm S}^{(j)}$.

\section{The exact model}\label{appendix-exact-dynamics}
In this section, following a procedure similar to \cite{rivas2010markovian}, we discuss  how to explicitly solve the exact dynamics of the Hamiltonian model for the joint system~${\cal S} + {\cal E}$.

Passing to the canonical variables  of the full model, i.e. introducing the operators
$x_{\rm A}=(a+a^\dag)/\sqrt{2}$, $p_{\rm A}=(a-a^\dag)/(\sqrt{2} i)$,
$x_{\rm B}=(b+b^\dag)/\sqrt{2}$, $p_{\rm B}=(b-b^\dag)/(\sqrt{2} i)$ as in Eq.~(\ref{eq:xA...pB}) and
$x_k=(c_k+c_k^\dag)/\sqrt{2}$, $p_k=(c_k-c_k^\dag)/(\sqrt{2} i)$,
the Hamiltonian (\ref{hamiltonian-decomposition}) of $\mathcal{S}+{\cal E}$ can be written as 
\begin{equation}
H=\frac{1}{2} \pmb{r}^T \mathcal{H} \pmb{r} + {\rm const}\,.
\end{equation}
The vector operator $\pmb{r}$ is the generalization of $\pmb{r}_{\rm S}$ introduced in Sec.~\ref{covariancesection}
that now contains the canonical coordinates of all the ${\cal S} + {\cal E}$ modes, i.e. 
\begin{eqnarray}
\pmb{r}&=&(x_{\rm A}, p_{\rm A}, x_{\rm B}, p_{\rm B}, x_1, p_1, ... , x_{M}, p_{M})^T\;, 
\end{eqnarray}
and $\mathcal{H}$ is a real symmetric $(2 M + 4 )\times (2 M + 4 )$  matrix, having non null elements only on the diagonal and on the first two  rows and on the first two  columns. This is because only the sub-system A is microscopically attached to the thermal bath:
\begin{equation}
\mathcal{H}=
\begin{pmatrix}
\omega_{\rm A} & 0 & g & 0 & \gamma_1 & 0 & \hdots & \gamma_{M} & 0\\
0 & \omega_{\rm A} & 0 & g & 0 &  \gamma_1& \hdots & 0 & \gamma_{M}\\
g & 0 & \omega_{\rm B} & 0 & 0 & 0 & \hdots & 0 & 0\\
0 & g & 0 & \omega_{\rm B} & 0 & 0 & \hdots & 0 & 0\\
\gamma_1 & 0 & 0 & 0 & \omega_1 & 0 & \hdots & 0 & 0\\
0 & \gamma_1 & 0 & 0 & 0 & \omega_1 & \hdots & 0 & 0\\
\vdots & \vdots & \vdots & \vdots & \vdots & \vdots & \ddots & \vdots& \vdots \\
\gamma_{M} & 0 & 0 & 0 & 0 & 0 & \hdots & \omega_{M} & 0\\
0 & \gamma_{M} & 0 & 0 & 0 & 0 & \hdots & 0 & \omega_{M} \\
\end{pmatrix}~.
\end{equation}
Exploiting the above construction the expectation value of $\pmb{r}$ can now be shown to  evolve in time as \cite{serafini2017quantum}
\begin{eqnarray}
\label{eq:first-mom-ev}
\langle \pmb{r}(t) \rangle : = \mbox{Tr}[ \pmb{r} \rho_{\rm SE}(t)] =e^{\Omega \mathcal{H}t} \langle \pmb{ r}(0) \rangle\;,
\end{eqnarray}
where $\Omega$ is the symplectic form of the entire model, i.e. the $(2 M + 4 )\times (2 M + 4 )$ matrix 
\begin{eqnarray} \label{COV} \Omega:= \bigoplus_{i=1}^{M+2} {\begin{pmatrix}
0 & 1\\
-1 & 0
\end{pmatrix}}\;, \end{eqnarray} 
whose elements embody  the canonical  commutation rules of entire ${\cal S} + {\cal E}$ system via the identity 
$\langle [\pmb{r}_\alpha,\pmb{r}_\beta]_- \rangle =i \Omega_{\alpha \beta}$.
Similarly  the covariance matrix of elements 
\begin{eqnarray}
\Sigma_{\alpha \beta}(t)&:=& \mbox{Tr}\left[ \Big[ \pmb{r}_\alpha - \langle \pmb{r}_\alpha(t) \rangle , 
\pmb{r}_\beta - \langle \pmb{r}_\beta(t) \rangle \Big]_+ \rho_{\rm SE}(t)\right]  
  \nonumber  \\ \label{eq:cov-mat-def}&=&
\left\langle \Big[ \pmb{r}_\alpha(t) - \langle \pmb{r}_\alpha(t)\rangle , 
\pmb{r}_\beta(t) - \langle \pmb{r}_\beta(t)\rangle\Big]_+\right\rangle,
\end{eqnarray}
can be shown to evolve as 
\begin{equation}
\label{eq:cov-mat-ev}
\Sigma(t)= e^{\Omega \mathcal{H}t} \Sigma(0) e^{ \mathcal{H} \Omega^Tt}\;.
\end{equation}
For future reference it is worth stressing  that the $4 \times 4$ principal minor of the  matrix $\Sigma(t)$ 
(i.e. the sub-matrix obtained from the latter by taking the upper left $4 \times 4$ part) 
corresponds to the covariance matrix $\Sigma_{\rm S}(t)$  of the ${\cal S}$  system alone, 
whose elements can be formally expressed as in Eq.~(\ref{COVAB}).

In the evaluation of    Eqs.~(\ref{eq:first-mom-ev}), (\ref{eq:cov-mat-ev}) one can  resort to the exact diagonalization of the Hermitian matrix $\mathcal{M}$
defined as
\begin{eqnarray}
\label{eq:i-omega-h}
\mathcal{M}:= i \Omega \mathcal{H}~.
\end{eqnarray}
Calling $(g_1, \hdots, g_{2M + 4})$ the eigenvalues of $\mathcal{M}$ and
\begin{eqnarray}
V_{\alpha \beta}:=[\pmb{\mathsf{g}}^{(\beta)}]_{\alpha}
\end{eqnarray}
the unitary matrix whose columns are the normalized eigenvectors $\pmb{\mathsf{g}}^{(\alpha)}$  corresponding to the eigenvalues 
$g_\alpha$, 
the diagonal form of the matrix $\mathcal{M}$ is obtained as:
\begin{eqnarray}
\mathrm{diag}(g_1, \hdots, g_{2M + 4})=
V^\dag \mathcal{M} V~. 
\end{eqnarray} 
Accordingly, we can now  rewrite Eqs.~(\ref{eq:first-mom-ev}), (\ref{eq:cov-mat-ev}) in the form
\begin{eqnarray}
\langle \pmb{r}(t)\rangle &=& V E_-(t) V^\dag \langle \pmb{r}(0)\rangle \nonumber\\
\Sigma(t) &=& V E_-(t) V^\dag \Sigma(0) V E_+(t) V^\dag~,
\label{eq:sigma-computational}
\end{eqnarray}
with 
\begin{eqnarray}
E_{\mp}(t)=\mathrm{diag}\left(e^{\mp i g_1 t}, \hdots, e^{\mp i g_{2M + 4} t} \right)~.
\end{eqnarray}
In summary, the exact dynamics is obtained thanks to the numerical diagonalization of the matrix
$\mathcal{M}$ of Eq.~(\ref{eq:i-omega-h}) and by performing the matrix multiplications in Eq.~(\ref{eq:sigma-computational}). 
Regarding the initial conditions, we observe that in the case of the input state we have selected in Eqs.~(\ref{ININ}), (\ref{iniENV}) and 
 (\ref{eq:in-cond-system}), the initial covariance matrix reads as the direct sum
\begin{equation}
\Sigma(0)=
\begin{pmatrix}
\mathds{1}_2 & \pmb{0} & \pmb{0}& \hdots& \pmb{0}\\
\pmb{0} & \mathds{1}_2 & \pmb{0} & \hdots& \pmb{0}\\
\pmb{0}& \pmb{0} & [2 \mathcal{N}(\omega_1)+1]\mathds{1}_2 & \hdots & \pmb{0}\\
\vdots & \vdots & \vdots & \ddots& \vdots \\
\pmb{0}& \pmb{0} &  \pmb{0} & \hdots & [2 \mathcal{N}(\omega_{M}) + 1]\mathds{1}_2
\end{pmatrix}
\end{equation}
with $\mathds{1}_2$ being the $2\times 2$ identity matrix and $\mathcal{N}(\omega_{k})$ being the Bose-Einstein mean occupation numbers
introduced in~Eq.~(\ref{eq:bose-factor}).
Regarding the first order moments instead, since  $\langle \pmb{r}(0) \rangle=0$ the evolution law of  Eq.~(\ref{eq:first-mom-ev}) leads to $\langle \pmb{r}(t) \rangle=0$ for all $t\geq 0$. 

\subsection{Memory and recurrence time scales}
\begin{figure*}
\vspace{.5cm}
\begin{overpic}[width=0.49\linewidth]{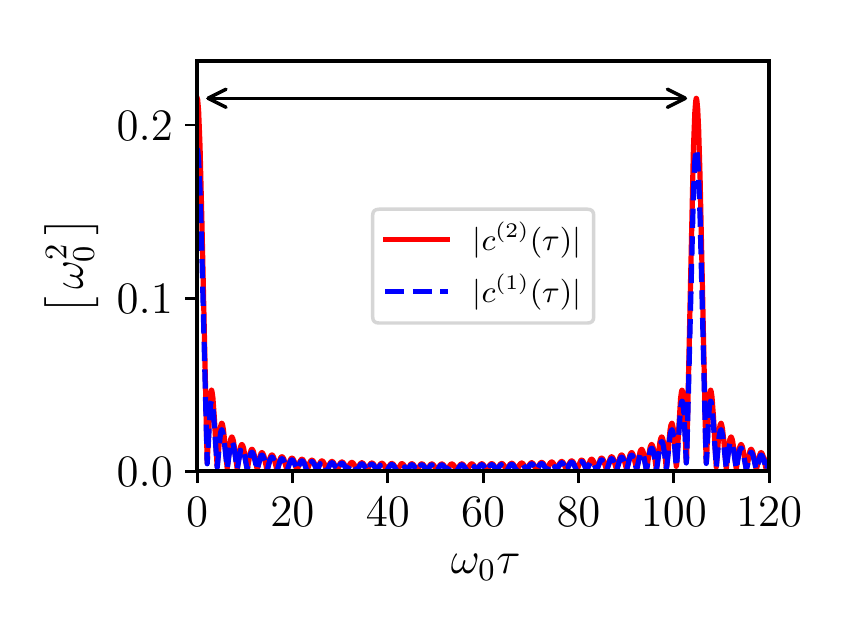}\put(5,75){(a)}\end{overpic}
\begin{overpic}[width=0.49\linewidth]{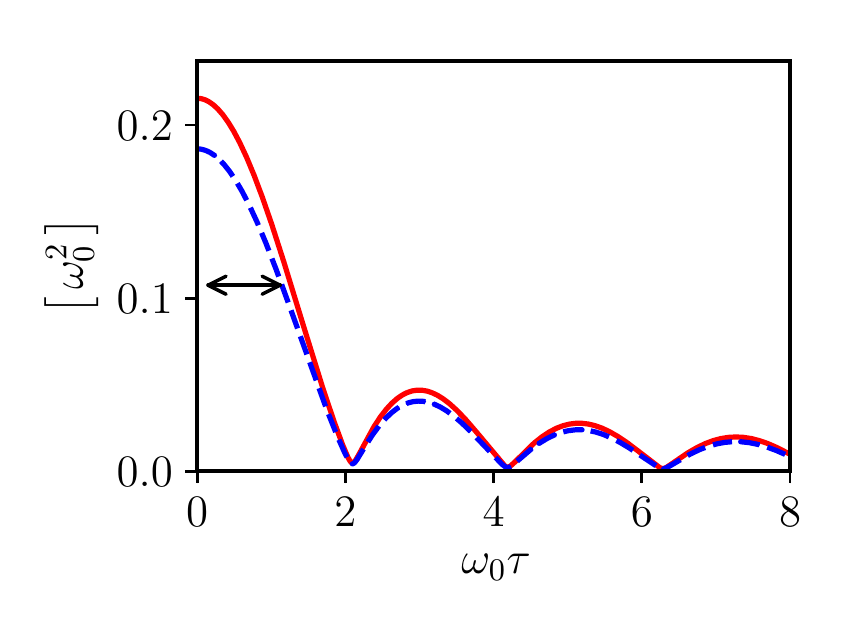}\put(5,75){(b)}\end{overpic}
\caption{\label{fig:rectime}
(Color online) Plot of the modulus of the bath correlation functions $c^{(1)}(\tau)$ and $c^{(2)}(\tau)$ 
(units $\omega_0^2$) 
defined in Eq.~(\ref{eq:c1c2}) that provide estimations  of the recurrence time (a) and  of the memory time (b). In Panel (a) we take $M=50$ oscillators in the thermal bath. We chose the parameters $\mathcal{N}(\omega_0)=10$,  $\kappa(\omega_0)=0.04 \omega_0$,  $\omega_c=3 \omega_0$ and $\alpha =1~.$}
\end{figure*}

\begin{figure*}
\begin{overpic}[width=0.49\linewidth]{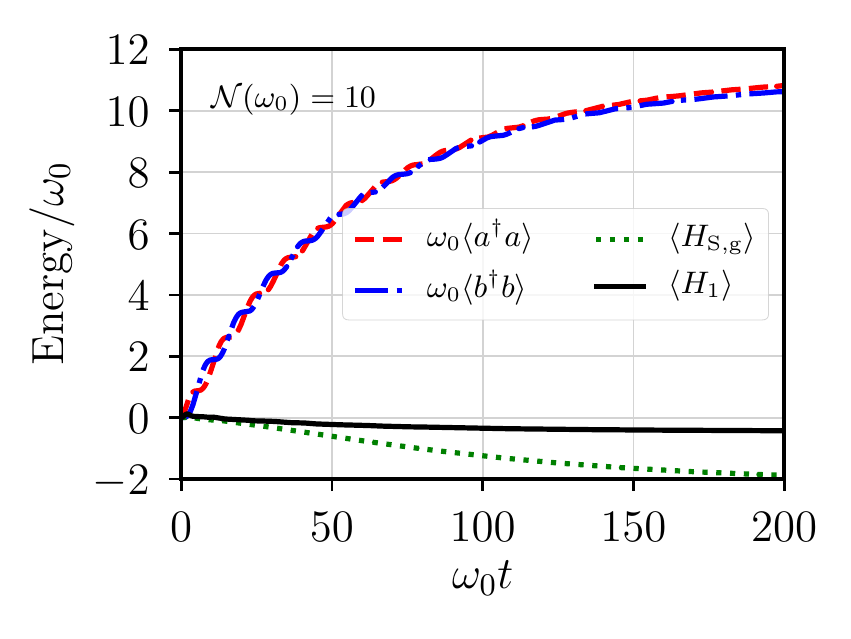}\put(5,75){(a)}\end{overpic}
\begin{overpic}[width=0.49\linewidth]{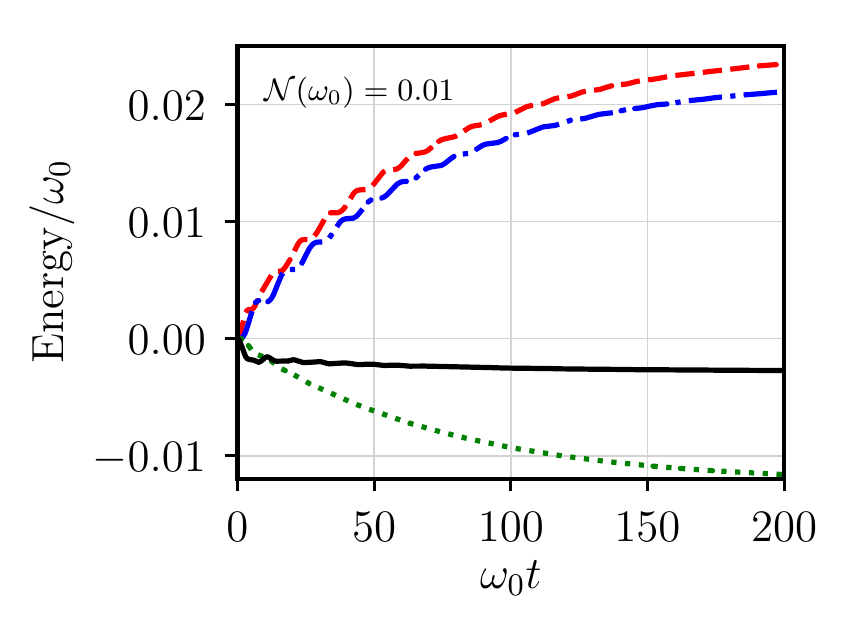}\put(10,75){(b)}\end{overpic}\\
\begin{overpic}[width=0.49\linewidth]{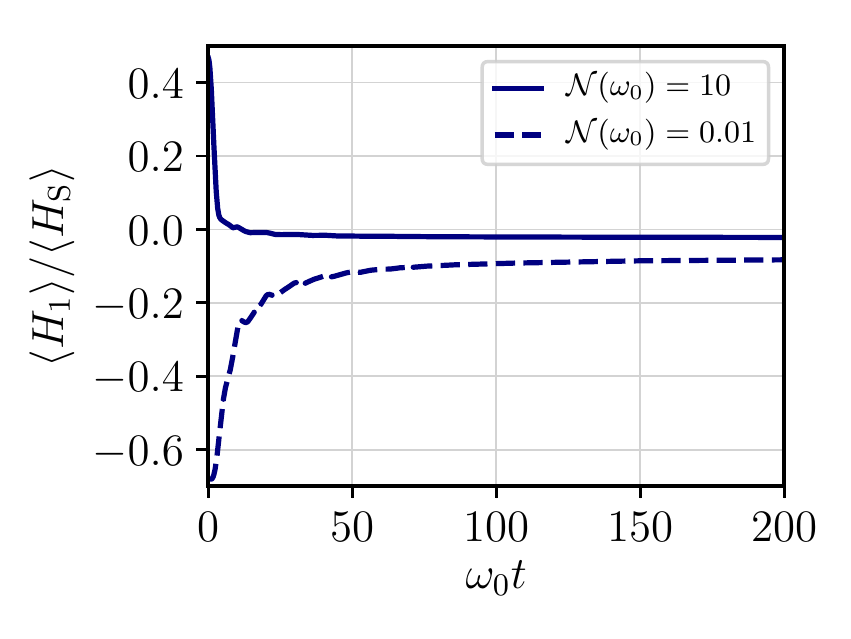}\put(5,70){(c)}\end{overpic}
\caption{\label{fig:energies}
(Color online) Time evolution of 
 the average components of the Hamiltonian~(\ref{hamiltonian-decomposition}) 
 obtained by numerically solving the exact dynamics of the full ${\cal S}+{\cal E}$ model in the high temperature regime $\mathcal{N}(\omega_0)=10$ (a), and in the low
temperature regime $\mathcal{N}(\omega_0)=0.01$  (b).
As indicated by the legend  the red dashed line corresponds to the local energy
of mode A; the blue dot-dashed line to the local energy term of mode B;
the green dotted line to the Hamiltonian A-B coupling term; and finally
 the black full line to the Hamiltonian ${\cal S} - {\cal E}$ coupling term. Notice that as the temperature
 decreases the incidence of the system-environment coupling gets relatively more consistent:
 this is explicitly shown in panel (c) where we report the ratio $\langle H_1 \rangle/\langle H_{\rm S} \rangle$
 for the two regimes.
In all the plots we assumed $g=0.3 \omega_0$, $\kappa(\omega_0)=0.04 \omega_0$, $\omega_c=3 \omega_0$, and $\alpha =1$.
}
\end{figure*}

When resorting to numerical methods in solving the exact Hamiltonian model one should be aware of the fact that since it
 involves a finite number of parties (i.e. the system modes A and B and
the $M$ environmental modes), it will be characterized by a recurrence time scale $T_{\rm rec}$ that, due to the various approximation involved in their
derivation, leave no trace in the corresponding ME expressions. 
An estimation of such quantity   can be retrieved directly from the periodicity of the correlation functions  of Eq.~(\ref{eq:c1c2}) which leads
us to (see Fig.~\ref{fig:rectime}a):
\begin{equation}
T_{\rm rec}=2 \pi M /\omega_c~.
\end{equation}
The choice of the parameters $\omega_c=3 \omega_0$ and $M\approx 400$ \cite{hofer2017markovian} ensures that the discretization  does not play any role in the time window we have considered for all the plots.

The width of the correlation functions~(\ref{eq:c1c2})  also  plays an important role in the model: 
it yields the time  $\tau_{\rm E}$ which takes for the information that emerges from the system 
to get lost into the environment and never coming back~\cite{breuer2002theory}.
Such time scale can't be resolved by any approximation we have discussed so far, because of the Markovian assumption which is present in all of them.
The estimation of this time scale is given by the half width at half maximum (see Fig.~\ref{fig:rectime}b) of  $|c^{(1)}(\tau)|$ and $|c^{(2)}(\tau)|$. 
For  $\mathcal{N}(\omega_0)=10$ we get 
\begin{equation}
\tau_{\rm E}\approx 3.8 / \omega_c~.
\end{equation}

\subsection{Low temperature effects} \label{LOWT} 

It is well known that  in the low temperature regime 
correlation effects between the bath and the system tent to arise,  
challenging the 
Born approximation used in the derivation of the Markovian MEs
~\cite{hovhannisyan2020charging}.
An evidence of this fact is presented in  Fig.~\ref{fig:energies}
where the time evolution of 
 the average components of the Hamiltonian~(\ref{hamiltonian-decomposition}) 
 are presented for two different choices of the parameter~$1/\beta$.


\section{On the thermalization of the system eigenmodes}
\label{appendix-eigenmodes}\label{APPEX} 
\begin{figure*}
\begin{overpic}[width=0.24\linewidth]{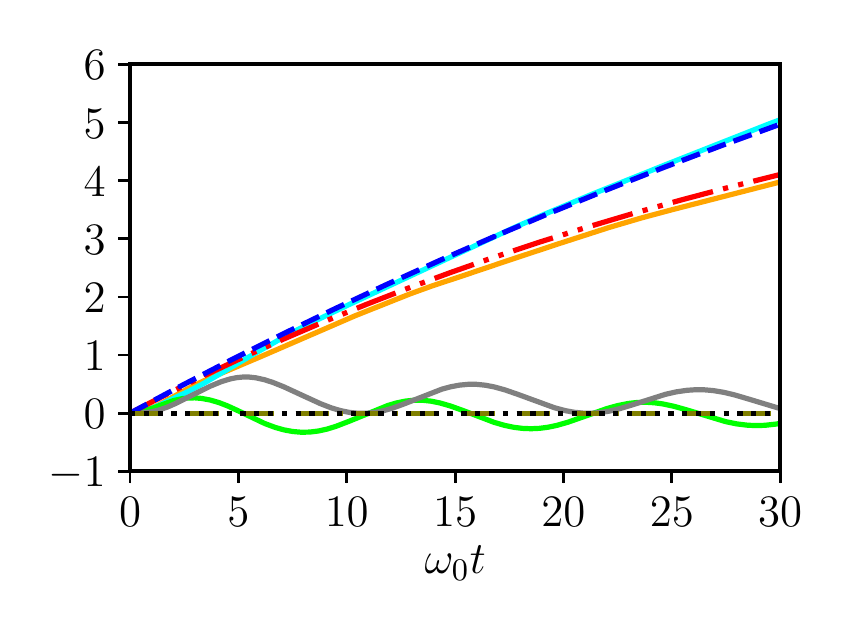}\put(30,73){(a)}\end{overpic}
\begin{overpic}[width=0.24\linewidth]{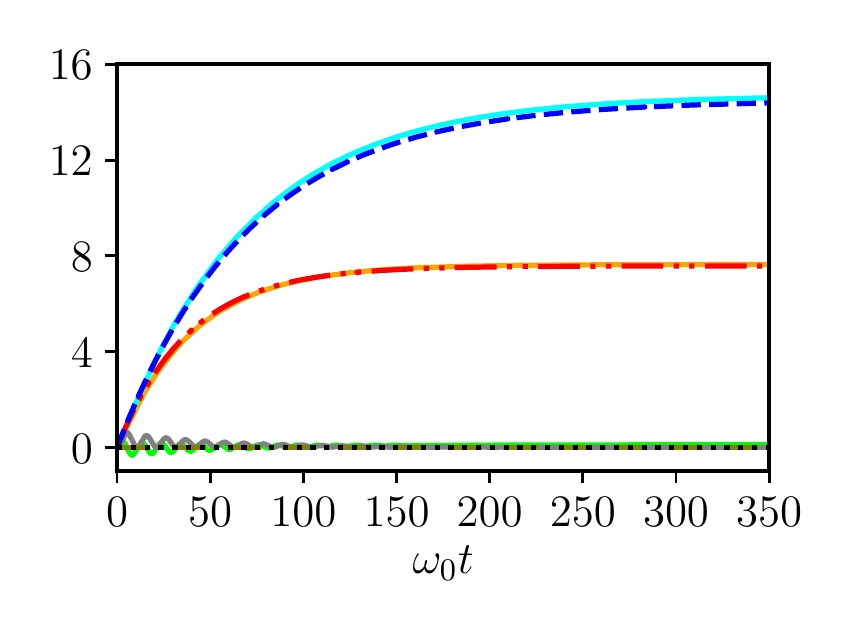}\put(30,73){}\put(-8,75){global}\end{overpic}
\begin{overpic}[width=0.24\linewidth]{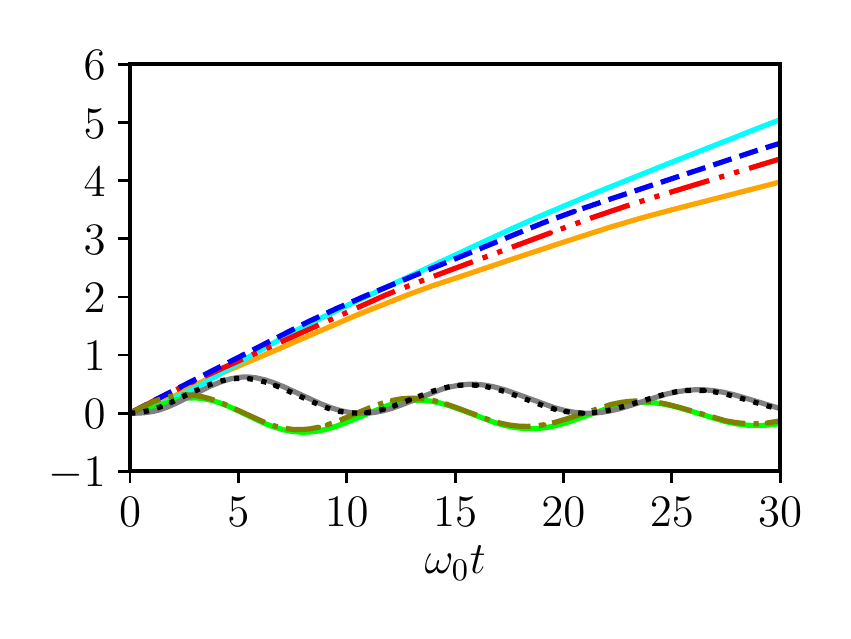}\put(30,73){(b)}\end{overpic}
\begin{overpic}[width=0.24\linewidth]{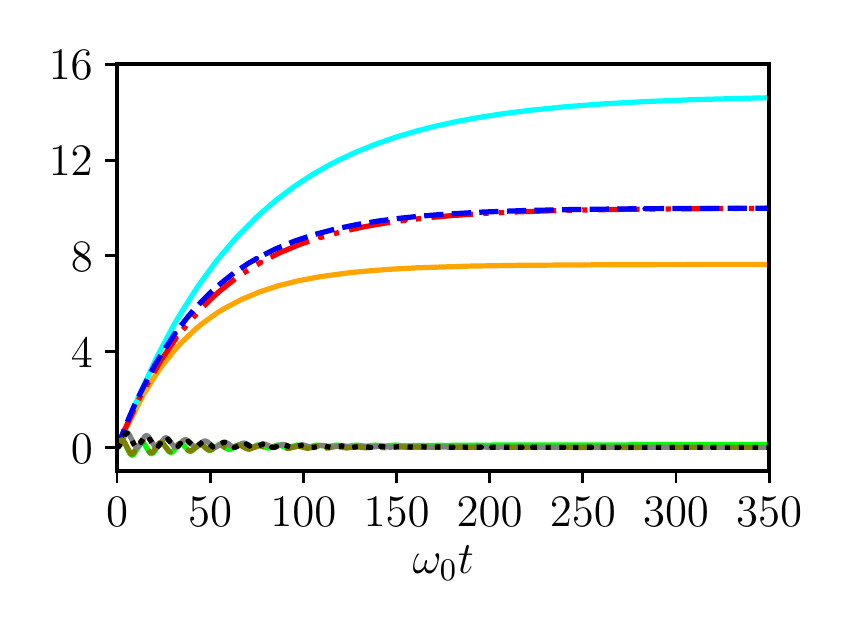}\put(30,73){}\put(-8,75){local}\end{overpic}
\\ \vspace{.5cm}
\begin{overpic}[width=0.24\linewidth]{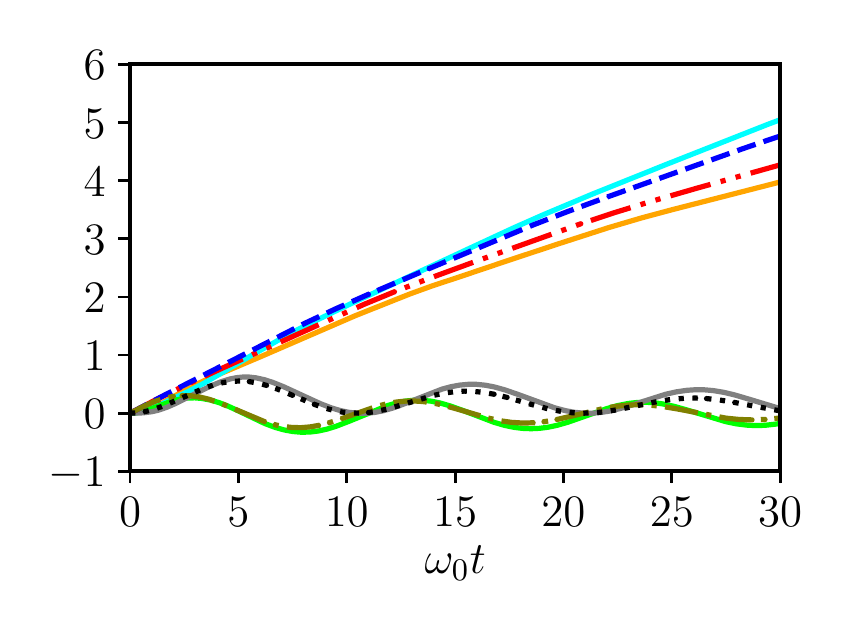}\put(30,73){(c)}\end{overpic}
\begin{overpic}[width=0.24\linewidth]{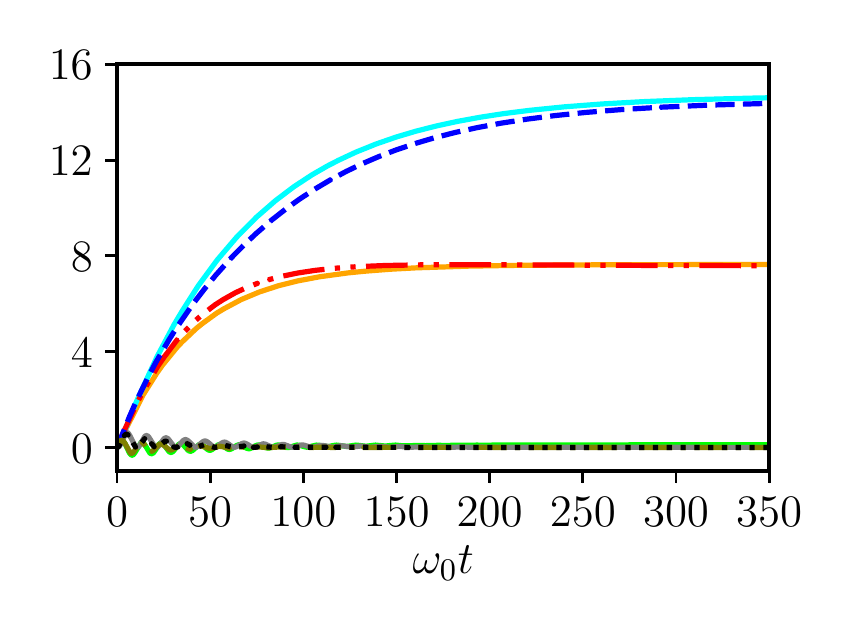}\put(30,73){}\put(-25,75){convex mixture}\end{overpic}
%
\begin{overpic}[width=0.24\linewidth]{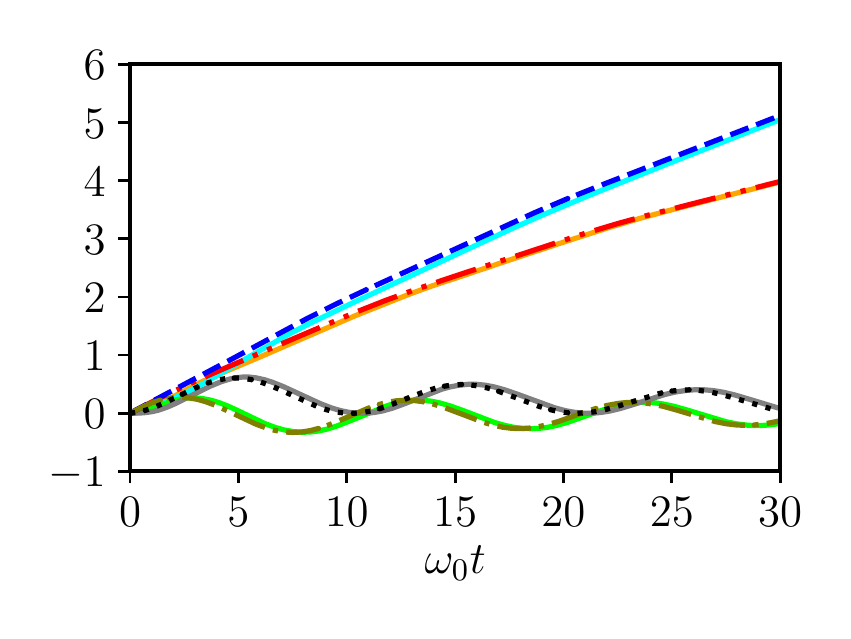}\put(30,73){(d)}\end{overpic}
\begin{overpic}[width=0.24\linewidth]{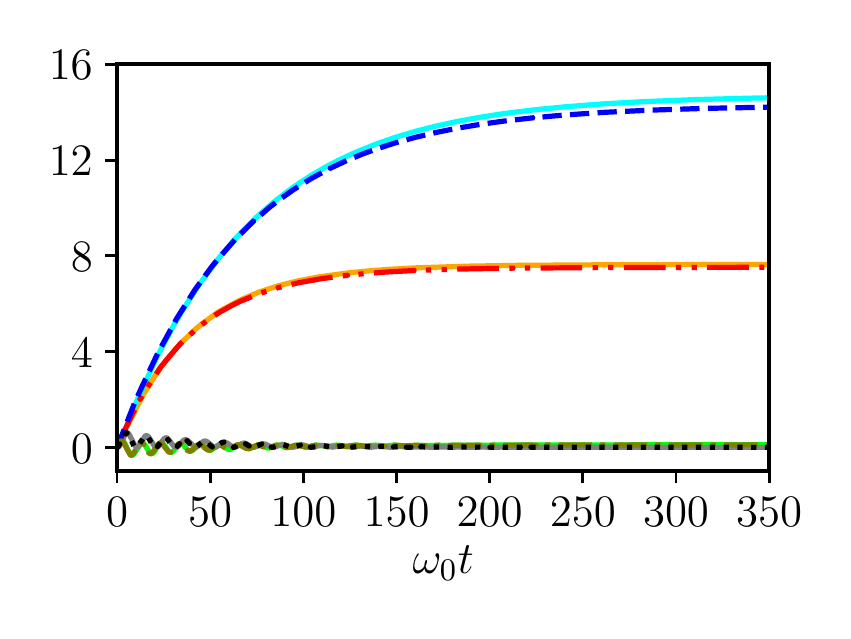}\put(30,73){}\put(-25,75){Redfield}\end{overpic}
\\ \vspace{.5cm}
\begin{overpic}[width=0.24\linewidth]{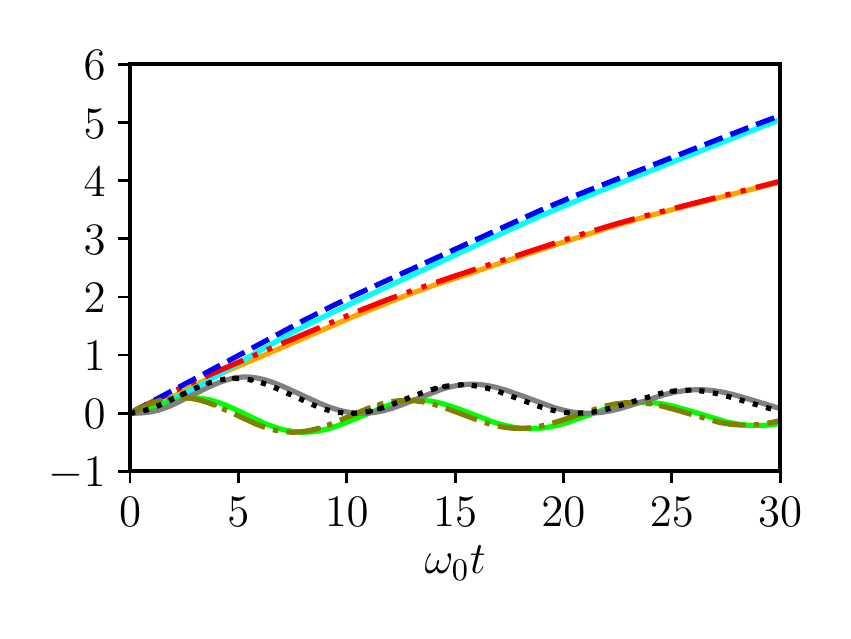}\put(30,73){(e)}\end{overpic}
\begin{overpic}[width=0.24\linewidth]{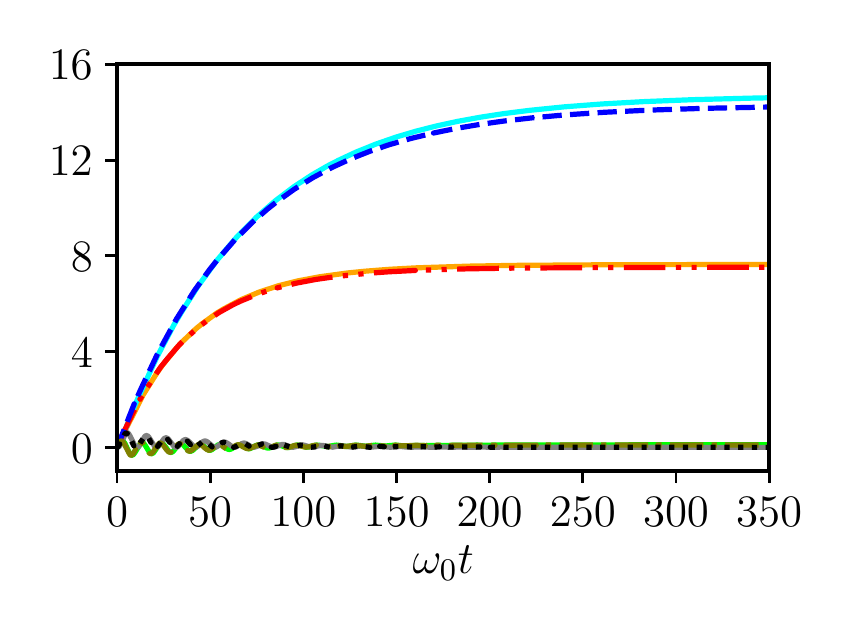}\put(30,73){}\put(-25,75){CP-Redfield}\end{overpic}
\begin{overpic}[width=0.38\linewidth]{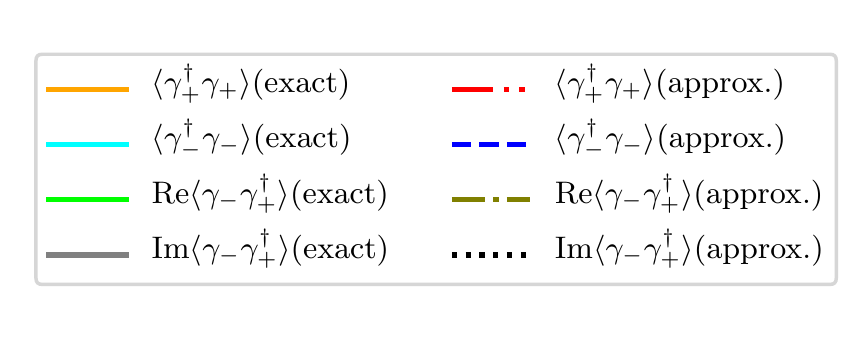}\end{overpic}
\caption{\label{fig:moms-pm}
(Color online) 
Comparison of second order moments in the eigenmodes basis evaluated using the global (a), local (b), convex mixture (c), Redfield (d), CP-Redfield (e) approximations with the ones predicted by the exact dynamics. 
As indicated by the legend continuous lines in the plots represent
the quantities computed by solving the exact ${\cal S} + {\cal E}$ Hamiltonian
model~(\ref{hamiltonian-decomposition}); dotted and dashed lines instead
refer to the approximated solutions.
Each panel contains two plots corresponding each to shorter (left) and longer (right) time scales.
We chose the parameters $\mathcal{N}(\omega_0)=10$,  $g=0.3 \omega_0$,  $\kappa(\omega_0)=0.04 \omega_0$,  $\omega_c=3 \omega_0$ and $\alpha =1~.$}
\end{figure*}

We show here the dual counterparts of the moments reported in Figs.~\ref{fig:moms-glob-loc} and \ref{fig:moms-redf} in the basis of the eigenmodes (\ref{eq:eigenmodes}),
making clearer when these eigenmodes reach the correct thermalization or not depending on the implemented approximation.
The second order moments in the  $a$, $b$ basis and the ones in the  $\gamma_+$, $\gamma_-$ basis are related each other as 
\begin{eqnarray}
\label{basis-change-1}
\frac{1}{2}( \langle a^\dag a\rangle-\langle b^\dag b\rangle)
&=& {\rm Re} \langle \gamma_- \gamma_+^\dag \rangle\;,\\
{\rm Im } \langle a b^\dag \rangle &=& {\rm Im } \langle \gamma_- \gamma_+^\dag \rangle ~,\\
{\rm Re } \langle a b^\dag \rangle &=& 
\frac{1}{2}( \langle \gamma_+^\dag \gamma_+\rangle-\langle \gamma_-^\dag \gamma_- \rangle)~,
\label{eq:pdp-minus-mdm}\\
\langle a^\dag a\rangle+\langle b^\dag b\rangle&=&\langle \gamma_+^\dag \gamma_+\rangle+\langle \gamma_-^\dag \gamma_- \rangle~.
\label{basis-change-4}
\end{eqnarray}
The steady state (\ref{eq:gibbsian-HS}) is what one expects from thermodynamics. It implies
$
\langle \gamma^\dag_{\rm \pm} \gamma_{\rm \pm} \rangle (\infty)= \mathcal{N(\omega_\pm)}~, 
\langle \gamma_{\rm -} \gamma^\dag_{\rm +} \rangle (\infty) =0~.    
$
This result is captured by applying the global approximation (see Eqs.~(\ref{eqs-GLOBAL})), which under the initial conditions (\ref{in-cond-p-m}) gives 
\begin{eqnarray}
\label{eq:analytic-global-pm}
\langle \gamma_{-} \gamma_{+}^\dag \rangle\Big|_{(\rm glob)}(t)&=&0,\\ 
\langle \gamma_{\pm}^\dag \gamma_{\pm} \rangle\Big|_{(\rm glob)}(t)&=&
\mathcal{N}(\omega_{\pm})\left(1-e^{-\frac{1}{2} \kappa (\omega_\pm) t}\right)~.
\end{eqnarray} 
On the other hand, the local approximation fails just about the steady state properties.
As discussed in~\cite{farina2019charger}, under the same initial conditions and when the Lamb-shift correction $\delta \omega_{\rm A}$ can be neglected,  the local ME (see Eqs.~(\ref{loc-eqs-pm})) leads to
\begin{eqnarray}
&& {\rm Re}\langle \gamma_- \gamma_+^\dag  \rangle\Big|_{(\rm loc)}
(t)=  \mathcal{N}(\omega_0) \frac{e^{-  \kappa(\omega_0) t/2}}{\epsilon }  \kappa(\omega_0)   \sin(\epsilon t/2)  
    ~,\nonumber \\
&&{\rm Im} \langle  \gamma_- \gamma_+^\dag \rangle\Big|_{(\rm loc)}(t)=
4 \mathcal{N}(\omega_0) \kappa(\omega_0) g
\frac{e^{-\kappa(\omega_0) t/2}}{\epsilon^2} 
[1-\cos(\epsilon t / 2)]~, \nonumber \\
&&\langle \gamma_{\pm}^\dag \gamma_{\pm} \rangle\Big|_{(\rm loc)}(t)
= \mathcal{N}(\omega_0) \nonumber \\
&&\qquad \times 
\{ 1-\frac{e^{-  \kappa(\omega_0) t/2}}{\epsilon^2}
    \left[16 g^2-\kappa(\omega_0)^2 \cos(\epsilon t/2)\right]\} 
\nonumber
~,
\end{eqnarray}
with 
$\epsilon:= \sqrt{(4 g)^2-\kappa(\omega_0)^2}$.
Using the relations (\ref{basis-change-1}-\ref{basis-change-4}), the above equations imply in the $a, b$ basis:
\begin{eqnarray}   
   && \langle a^\dag a \rangle\Big|_{(\rm loc)}
(t)=  \mathcal{N}(\omega_0) \{ 1-\frac{e^{-  \kappa(\omega_0) t/2}}{\epsilon^2}\nonumber\\ &&\qquad
   \quad  \times \left[16 g^2-\kappa(\omega_0) \epsilon \sin(\epsilon t/2)-\kappa(\omega_0)^2 \cos(\epsilon t/2)\right]\} 
    ~,\nonumber \\
&&   \langle b^\dag b \rangle\Big|_{(\rm loc)}(t)= 
      \mathcal{N}(\omega_0) \{1-\frac{e^{-  \kappa(\omega_0) t/2}}{\epsilon^2}\nonumber\\ &&\qquad
     \quad  \times  \left[16 g^2+\kappa(\omega_0) \epsilon \sin(\epsilon t/2)-\kappa(\omega_0)^2 \cos(\epsilon t/2)\right]\} 
~,\nonumber \\
&&{\rm Im} \langle  a b^\dag \rangle\Big|_{(\rm loc)}(t)=
4 \mathcal{N}(\omega_0) \kappa(\omega_0) g
\frac{e^{-\kappa(\omega_0) t/2}}{\epsilon^2} 
[1-\cos(\epsilon t / 2)]~, \nonumber 
\end{eqnarray} 
\begin{eqnarray} 
{\rm Re}\langle  a b^\dag \rangle\Big|_{(\rm loc)}(t)=0~,
\nonumber \end{eqnarray}
i.e.
\begin{eqnarray}
&&\langle H_{\rm S,g} \rangle\Big|_{(\rm loc)}(t)=0 ~.
\end{eqnarray}
In Fig.~\ref{fig:moms-pm}
we plot the moments in the eigenmodes basis, comparing the results obtained by the Global, Local, convex mixture, Redfield, CP-Redfield approximations with the ones predicted by the exact dynamics, by including this time also the Lamb-shift contributions. In the local case for instance the Lambshift implies a tiny splitting between $\langle \gamma_+^\dag \gamma_+\rangle\Big|_{(\rm loc)}(t)$ and $\langle \gamma_-^\dag \gamma_-\rangle\Big|_{(\rm loc)}(t)$ at short time scales (connected to a small but non-vanishing ${\rm Re} \langle a b^\dag \rangle\Big|_{(\rm loc)}(t)$,  see Eq.~(\ref{eq:pdp-minus-mdm})).
Again the convex mixture of the local and global approximations of Eq.~(\ref{eq:convex-comb}) and the CP-Redfield equation yield a very good approximation either of the transient than of the steady state properties.

\clearpage

\vspace{3cm}
%



\end{document}